\documentclass[fleqn,usenatbib]{mnras}
\usepackage{newtxtext,newtxmath}
\usepackage[normalem]{ulem}
\usepackage{tabularx}
\newcolumntype{Y}{>{\raggedright\arraybackslash}X}  
\usepackage[T1]{fontenc}

\DeclareRobustCommand{\VAN}[3]{#2}
\let\VANthebibliography\thebibliography
\def\thebibliography{\DeclareRobustCommand{\VAN}[3]{##3}\VANthebibliography}

\usepackage{graphicx}	
\usepackage{amsmath}	
\usepackage{todonotes}
\usepackage{microtype}
\usepackage{xcolor}
\usepackage{float}
\usepackage{comment}
\usepackage{xspace}

\def\HI{\ion{H}{I}\xspace}
\def\Hi{\ion{H}{I}}
\def\HII{\ion{H}{II}~}
\def\xHI{x_{\rm \ion{H}{I}}}
\def\xHII{x_{\rm \ion{H}{II}}}
\def\xb{\bar{x}_{\rm \ion{H}{I}}}
\def\xmb{\bar{x}_{\rm \ion{H}{Im}}}
\def\xa{\bar{x}_{\rm \ion{H}{II}}}
\def\xc{\bar{x}_{\rm \ion{H}{IIm}}}
\def\lya{Ly$\alpha$\xspace}
\defcitealias{Xu2019}{X19}
\defcitealias{Worseck2014}{W14}	
\defcitealias{Bosman2022}{B22}

\usepackage{lipsum} 
\usepackage{comment}

\usepackage{tikz,xcolor,hyperref}
\definecolor{lime}{HTML}{A6CE39}
\DeclareRobustCommand{\orcidicon}{%
	\begin{tikzpicture}[baseline=-0.8ex]
	\draw[lime, fill=lime] (0,0) 
	circle [radius=0.16] 
	node[white] {{\fontfamily{qag}\selectfont \tiny ID}};
	\draw[white, fill=white] (-0.0625,0.095) 
	circle [radius=0.007];
	\end{tikzpicture}
	\hspace{-2mm}
}

\foreach \x in {A, ..., Z}{%
	\expandafter\xdef\csname orcid\x\endcsname{\noexpand\href{https://orcid.org/\csname orcidauthor\x\endcsname}{\noexpand\orcidicon}}
}

\title[IGM parametrisation]{An Intergalactic medium centred parametrisation of the 21-cm power spectrum from the Epoch of Reionization}

\author[I. Georgiev et al.]{
Ivelin Georgiev$^{1,2}$\thanks{E-mail: igeorgiev@mpa-garching.mpg.de}\orcidA,
Saleem Zaroubi$^{3,2}$\orcidB,
Raghunath Ghara$^{4}$\orcidC,
Abinash Kumar Shaw$^{1,2}$\orcidD,
\newauthor
Garrelt Mellema$^{5}$\orcidE,
Benedetta Ciardi$^{1}$\orcidF,
L. V. E. Koopmans$^{3}$\orcidG,
Anshuman Acharya$^{6}$\orcidH,
\newauthor
Sambit K. Giri$^{5}$\orcidI,
Ilian T. Iliev$^{7}$\orcidJ,
and Qing-bo Ma$^{8,9}$\orcidK
\\
$^{1}$Max-Planck-Institut für Astrophysik, Karl-Schwarzschild-Straße 1, 85748 Garching, Germany\\
$^{2}$ARCO (Astrophysics Research Center), Department of Natural Sciences, The Open University of Israel, 1 University Road, PO Box 808, Ra'anana 4353701, Israel\\
$^{3}$Kapteyn Astronomical Institute, University of Groningen, PO Box 800, 9700 AV Groningen, The Netherlands\\
$^{4}$Department of Physics, Indian Institute of Technology Kharagpur, Kharagpur, West Bengal 721302, India\\
$^{5}$Department of Astronomy and Oskar Klein Centre, AlbaNova, Stockholm University, SE-10691 Stockholm, Sweden\\
$^{6}$Berkeley Center for Cosmological Physics (BCCP), University of California, Berkeley, CA 94720, United States\\
$^{7}$Astronomy Centre, Department of Physics \& Astronomy, Pevensey III Building, University of Sussex, Falmer, Brighton, BN1 9QH, United Kingdom\\
$^{8}$School of Physics and Electronic Science, Guizhou Normal University, Guiyang 550001, PR China\\
$^{9}$Guizhou Provincial Key Laboratory of Radio Astronomy and Data Processing, Guizhou Normal University, Guiyang 550001, PR China
}

\date{Submitted to MNRAS. Received YYY; in original form ZZZ}

\pubyear{2026}

\begin{document}
\label{firstpage}
\pagerange{\pageref{firstpage}--\pageref{lastpage}}
\maketitle
\begin{abstract}
The 21-cm signal from the intergalactic medium (IGM) is a key probe of the Epoch of Reionization (EoR). In this work, we present a model of the 21-cm power spectrum centred on the state of the IGM. The framework combines aspects of perturbation and percolation theory to deduce a physically informed parametrisation of the power spectrum. Moreover, it complements classical interpretation methods, centred on modelling the properties of the first ionising sources. We define the IGM parameters driving the shape and amplitude of the 21-cm signal, from the globally averaged volume- and mass-weighted ionisation fractions ($\xa$ and $\xc$), the clustering of sources, through the halo bias ($b_{\mathrm{h}}$), and the characteristic sizes of ionised regions ($R_{i}$) and neutral islands ($R_{n}$). We extend this analysis to model the redshift dependence of the power spectrum, constraining parameters such as the midpoint and duration of reionization. We test the IGM framework on three reionization simulations ({\sc C$^{2}$-Ray}, {\sc 21cmFAST}, and {\sc POLAR}) with varying cosmologies, astrophysics, and numerical assumptions. The framework constrains the IGM parameter space, recovers the midpoint and duration of the EoR as well as the minimum mass of halos hosting ionising sources $M_{\mathrm{min}}$, and the redshift evolution of the global ionisation fraction, the evolution of ionised and neutral regions (measured with the radial mean free path method by Georgiev et al. (2026b, submitted)). Moreover, the IGM framework lends itself to joint analysis with other EoR probes, such as high-redshift galaxies and quasars and the Cosmic Microwave Background.
\end{abstract}
\section{Introduction}
\label{sec:intro}
The first billion years of the Universe are a hallmark period, defined by the active formation and growth of the first galaxies. However, the first stars and active galactic nuclei not only affected the formation history of their host galaxies, but also actively shaped the state of the medium between them. The intergalactic medium (IGM), primarily consisting of neutral hydrogen ($\HI$), was re-ionised by the ultraviolet photons from these first ionising sources, leading to a large-scale phase transition we refer to as the Epoch of Reionization (EoR). The EoR is a remarkable process as it is driven by the collective astrophysical properties of the first ionising sources, which themselves are tracers of cosmic structure \citep{Zaroubi2013,mcquinn2016evolution,dayal2018early}.
A key probe of reionization comes from observations of the Cosmic Microwave Background (CMB), whose measurements of the Thomson optical depth from \citet{Planck2020} have constrained a value of $\tau =0.054 \pm 0.007$, corresponding to a midpoint of reionization of $z_{\mathrm{re}} \sim 7.5$ \citep[though see][]{Cain2025,Sailer2026}. Moreover, studies of secondary CMB anisotropies such as the patchy kinetic Sunyaev-Zel'dovich effect (pkSZ) \citep{Aghanim1996,Gruzinov1998} have placed a $2 \sigma$ upper limit on the duration of reionization of $\Delta z <4.1$ \citep{Reichardt2021,Chaubal2026}.  Another key parameter of the progress of reionization is the global volume-averaged neutral hydrogen fraction ($\xb$). This can be probed by studying the spectra of the ionising sources themselves, as emitted \lya photons will scatter as they travel across the neutral IGM during the EoR. Few examples of these are the observations of the Gunn-Peterson effect \citep{Gunn-Peterson1965} in quasar (QSO) spectra \citep{Fan2006,Mortlock2011,Greig2017,Banados2018,Greig2019,Wang2020,Yang2020,Becker2021} as well as studies of dark gaps \citep{Zhu2021,Zhu2022} and dark pixels \citep{Davies2026}. These studies have provided tight constraints on the final phases of reionization, estimating the end of the EoR to lie in the range $z_{\mathrm{end}}\approx 5-5.5$ \citep{Bosman2022}. On the other hand, QSO spectra observations have also been shown to be able to constrain the sizes of neutral patches ($R_n$), referred to as $\Hi$ islands \citep{Becker2015}. The advent of the James Webb Space Telescope (JWST) has pushed our understanding of the EoR even further \citep[see][]{naidu2022two,robertson2023identification,2023ApJ...949L..42M,donnan2024jwst,cameron2023jades}, allowing for measurements of the global neutral fraction by studying the \lya equivalent width \citep{tang2023jwst,Nakane2024,Jones2024}. Moreover, studies investigating the shape of the UV continuum redward of \lya, the \lya damping wing, have also placed constraints on $\xb$ and the sizes of ionised regions during the EoR ($R_i$) \citep{CurtisLake2023,Umeda2023,Mason2026}. Furthermore, UV luminosity functions from high-redshift galaxies measured with the Hubble Space Telescope and the JWST have been used to constrain EoR properties \citep[see][and references therein]{Bouwens2021,Whitler2025,Atek2026}. Note, however, that estimates from a selection of ionising sources are sensitive to biasing due to sample variance and the role of their local diversity \citep[see][for a discussion]{Lu2024}.

One of the most prominent prospective probes of the EoR comes from the state of the IGM itself. This is due to the nature of neutral hydrogen, which can either emit or absorb photons with a rest wavelength of $\lambda = 21.106$\,cm \citep{Wouthuysen1952,Field1957}, depending on its hyperfine structure and coupling to the temperature of the IGM. Conversely, ionised hydrogen does not emit a 21-cm signal. Therefore, the temperature and ionisation state of the IGM can be probed directly by the 21-cm signal across a wide range of redshifts and profoundly impact our understanding of cosmology and astrophysics. One detection method consists of measuring the sky-averaged 21-cm signal using single-dish antennae. Such global 21-cm experiments include SARAS \citep{Patra2015}, SARAS2 \citep{Singh2018}, LEDA \citep{Price2018}, and EDGES \citep{Bowman2010} $\&$ EDGES 2 \citep{Monsalve2017} as well as REACH \citep{deLeraAcedo2022}. While the EDGES team have claimed a detection of the signal \citep{Bowman2018}, its distinct and deep absorption profile has not been confirmed by the {\sc SARAS3} experiment \citep[see][for an overview]{Singh2022}.
Another prospect is detecting the fluctuations in the 21-cm signal, a key science goal of multiple low-frequency radio interferometers. Examples of such telescopes include the Low-Frequency Array (LOFAR)\footnote{\url{www.lofar.org}} \citep{vanHaarlem2013}, the New Extension in Nançay
Upgrading LOFAR (NenuFAR) \citep{Zarka2020}, the Hydrogen Epoch of Reionization Array (HERA)\footnote{\url{https://reionization.org}} \citep{DeBoer2017}, the Murchison Widefield Array (MWA)\footnote{\url{www.mwatelescope.org}} \citep{Bowman2013}, and the forthcoming Square Kilometre Array (SKA)\footnote{\url{www.skatelescope.org}} \citep{SKA2026}, whose increased sensitivity should also allow us to study tomographic data \citep{Mellema2015}. The power spectrum of the 21-cm signal is the first cornerstone for understanding the EoR \citep{Furlanetto2006a} and a key goal of the aforementioned radio interferometers. 

Although the signal has not yet been detected, multiple upper limits have been placed across a wide range of redshifts and wavenumbers, continually improving analysis methods and reducing uncertainties from foregrounds, radio frequency interference, and other systematics \citep{Liu2020}. Examples include the upper limits from LOFAR using the North Celestial Pole field \citep{Mertens2020,Mertens2025} as well as the recent progress on analysis of the 3C 196 field \citep{Ceccotti2025}, MWA \citep{Trott2020,Yoshiura2021,Nunhokee2025}, HERA \citep{HERA2021,Hera2026}, NenuFAR \citep{Munshi2025}, GMRT \citep{Paciga2013}, and PAPER \citep{Kolopanis2019}. Notably, such upper limits have begun constraining the parameter space of the properties of the first ionising sources \citep{ghara2020constraining,Mondal2020,Greig2021,Ghara2021,Ghara2025} and excluding exotic reionization models \citep[see][for a discussion]{HERA2022}.

The motivation behind this work centres on the philosophy of interpreting 21-cm power spectra and their upper limits. The aforementioned analyses are grounded in the properties of the first ionising sources. Generating a large set of reionization simulations across the parameter space of the first ionising sources, the 21-cm power spectra are a by-product that is then used for statistical inference \citep[e.g.][]{Greig2015,GreigMesinger2017}. While successful and practical, as it can be analysed considering high-$z$ direct observations of the first ionising sources, such a ``source-centred'' analysis will naturally have some disadvantages. First, such methods demand a large number of EoR simulations, requiring thousands, if not tens of thousands, of models to properly sample the high-dimensional parameter space \citep{Greig2015,Breitman2024}. This places a numerical uncertainty on the analysis, requiring the use of faster but less accurate semi-numerical models, which cannot account for full radiative transfer \citep{Zahn2011,Hutter2018}, non-linear clustering of sources \citep{Iliev2006b, Trac2007}, or baryonic effects seen in high-resolution (but small volume) hydrodynamic simulations \citep{Kannan2022}. This is exacerbated by the need to match the large field of view of the observations and a high enough resolution to properly capture the reionization morphology \citep{Iliev2014,Kaur2020}. Since we cannot resolve galaxies in such simulations, parametrisations of the first sources are required, which can be difficult to define and highly degenerate \citep{Park2019,Berklas2026}. On the other hand, analytically modelling the 21-cm power spectrum from first principles is arduous and requires a deep understanding of astrophysics and cosmology. Analytic models \citep{Furlanetto2004, Schneider2021, Schneider2023, Munoz2023} provide a solid foundation for how reionization progresses and evolves, yet such models fail to capture effects such as the overlapping of ionised structures and cannot reproduce the cumulative growth of the ionised regions \citep[see][for a discussion]{Lin2016}. Meanwhile, interpretations by \cite{Ghara2024} and \cite{Mirocha2022} show that the power spectrum can indeed be constrained by an analytic form with a set of parameters. However, these methods share a significant drawback. Because their functional form and parameters are phenomenological by design rather than tied to the physical quantities of the EoR, it is difficult to meaningfully constrain their prior ranges or model how they evolve with redshift, which limits their use for power spectrum interpretation.

The goal of this work is to derive and utilise a parametrisation of the 21-cm power spectrum grounded in IGM properties and driven by physically informed parameters that can be meaningfully constrained and used for statistical inference. Such an analysis is grounded in the fact that the 21-cm power spectrum originates from the neutral IGM, rather than the ionising sources themselves, hence, we choose to model their general impact on the IGM morphology. Therefore, such a parametrisation offers an approach to studying the 21-cm power spectrum, which is complementary to the current interpretation methods, and will allow us to further our understanding of reionization.

The paper is structured as follows. In Sec.~\ref{sec:theory} we explain the theory behind the 21-cm signal as well as the IGM framework we have developed. Section~\ref{sec:sims} showcases the different reionization simulations used to verify our method, while Sec.~\ref{sec:mcmc} introduces the Markov Chain Monte Carlo (MCMC) we use in the IGM framework. The results and constraints on the IGM properties are showcased for the different models in Sec.~\ref{sec:results}, followed by a summary and conclusions in Sec.~\ref{sec:conclusions}. The paper also includes additional information on the framework's theory in Appendix~\ref{sec:cross-terms} as well as a forecast study of the IGM framework in Appendix~\ref{sec:forecast}.
\section{Theory}
\label{sec:theory}
The 21-cm line is the product of the spin-flip transition of the ground state of neutral hydrogen \citep{Wouthuysen1952,Field1957}. The 21-cm signal detected by radio interferometers is expressed through the differential surface brightness temperature of the signal against that of the CMB. At low frequencies, in the Rayleigh-Jeans limit, we can express the differential surface brightness temperature as
\begin{equation}
\label{eq:Tb}
\delta T_{\mathrm{b}} (\mathbf{r}, z) =\frac{1-e^{-\tau(\mathbf{r}, z)}}{1+z}\big(T_{\mathrm{s}} (\mathbf{r}, z)-T_{\mathrm{CMB}}(z)\big)\,,
\end{equation}
where $\tau$ is the optical depth of the gas. 
The spin temperature $T_{\mathrm{s}}$ is defined by the level populations of the 21-cm transition and must differ from $T_\mathrm{CMB}$ in order to produce an observable signal. During reionization, active star formation in the first galaxies will result in the emission of \lya and X-ray radiation, which will increase the temperature of the IGM, coupling the spin temperature to the kinetic temperature \citep{Wouthuysen1952,Field1959,Pritchard2007}. 
In the optically thin regime ($\tau \ll 1$) we can express Eq.~\eqref{eq:Tb} in the following form \citep{Mellema2013} 
\begin{align}
\label{eq:TbS}
\delta T_{\mathrm{b}} (\mathbf{r}, z) = T_{0} (\mathbf{r},z) \xb(1+\delta_{x_{\HI}}(\mathbf{r}))(1+\delta_{\rho}(\mathbf{r})) \ ,\mathrm{where} 
\end{align}
\begin{align}
&T_0 (\mathbf{r},z) \approx 27 
\bigg(\frac{1+z}{10}\bigg)^{1/2} \bigg(\frac{T_{\mathrm{s}}(\mathbf{r},z) -T_{\mathrm{CMB}}(z) }{T_{\mathrm{s}}(\mathbf{r},z) }\bigg)  
\nonumber
\\ &\bigg( \frac{\Omega_{\mathrm{b}}}{0.044}  \frac{h}{0.7} \bigg) \bigg( \frac{\Omega_{\mathrm{m}}}{0.27}  \bigg)^{-1/2} 
\bigg( \frac{1-Y_{\mathrm{p}}}{1-0.248} \bigg) ~\mathrm{mK}.  
\end{align}
We express the fluctuation of the density of matter $\rho$ as $\delta_{\rho}= \rho/\bar{\rho} -1 $, while the fluctuation of the neutral fraction field $\xHI$ is $\delta_{x_{\HI}}$. Note that 
$T_{0}$ term encapsulates the dependence of the differential surface brightness temperature on the cosmological parameters and the spin temperature of the gas. Here, $\Omega_m$ and $\Omega_b$ are the critical density parameters of matter and baryons, $Y_p$ and $h$ are the primordial helium fraction and the dimensionless Hubble constant.

In this work, we consider the limit where the spin temperature is saturated and thus $T_{\mathrm{s}}\gg T_{\mathrm{CMB}}$. In doing so, we do not account for the fluctuations of $T_{\mathrm{s}}$, and the framework therefore applies to the EoR rather than to the earlier Cosmic Dawn. We will address these fluctuations and the additional cross terms they introduce \citep{Schaeffer2024} in future work. In this form, $T_{0}$ is scale-independent, hence, we will divide it out from our model and assume the background cosmology to be well constrained. In taking the Fourier transform of Eq.~\eqref{eq:TbS}, $P_{21cm} (k) = \langle \delta T_{\mathrm{b}} (\boldsymbol{k}),\delta T_{\mathrm{b}} ^{*} (\boldsymbol{k})\rangle$, we can express the autocorrelation power spectrum of the 21-cm field, and expand the fluctuations of the field in the following form
\begin{align}
\label{eq:Decomposition}
\frac{P_{21 \mathrm{cm}}}{T_{0}^{2}(z)} (k) = \xb^{2}\Big( P_{\delta_{\rho},\delta_{\rho}}(k)+P_{\delta_{x_{\HI}},\delta_{x_{\HI}}}(k) +2P_{\delta_{\rho},\delta_{x_{\HI}}} (k) \\ 
+2P_{\delta_{\rho}\delta_{x_{\HI}},\delta_{x_{\HI}}}(k) +2P_{\delta_{\rho}\delta_{x_{\HI}},\delta_{\rho}}(k) +P_{\delta_{\rho}\delta_{x_{\HI}},\delta_{\rho}\delta_{x_{\HI}}} (k) 
\Big) \nonumber.
\end{align}
The terms on the top line are the autocorrelation power spectra of the density and neutral fields, as well as their cross-correlation, and are key components that define the properties of the 21-cm power spectrum throughout the EoR. Conversely, the terms on the bottom line express cross-terms between multiples of the two fields, which are often non-trivial in form yet important for the overall shape and evolution of the 21-cm power spectrum \citep[see][for a detailed discussion]{Lidz2007,Georgiev2022}. We consider the role of each of these terms in shaping the 21-cm signal during the EoR in the following sections.
\subsection{Power spectrum IGM parametrisation}
\label{sec:21ps_parametrisation}
To derive the parametrisation of the 21-cm power spectrum shown later in Eq.~\eqref{eq:p21_model}, we first consider the general philosophy of our approach. Pivotal works such as \citet{Furlanetto2004} formulate this through the lens of the sources of reionization and their properties. In this work, we adopt a complementary approach, emphasising the 21-cm field and, by extension, the IGM itself. Rather than defining the properties of the first sources of ionisation, we model their general effect on the morphology of the IGM to devise a framework based on ``IGM'' parameters. 
\subsubsection{The large-scale power spectrum of the ionisation field through Standard Perturbation Theory}
\label{eq:SPT}
First, we simplify this task by considering the absolute ionisation fluctuation, which we will define as $\delta_{x} = \xHII - \xa$, as it is a bounded quantity. In this form, the autocorrelation power spectrum $P_{\delta_{x},\delta_{x}} = P_{\xHII,\xHII}$, which is one of the main terms guiding the evolution of the 21-cm signal \citep{Lidz2007}. We treat the ionisation/neutral field as a two-phase medium (and hence a bimodal distribution)\footnote{We neglect partial ionisations in the derivation. These can have both numerical and physical origins, and their effect on the fluctuations is non-trivial. We return to this in Sec.~\ref{sec:polar_results}.}. We firstly focus on the ionisation component, returning to the neutral component in Sec.~\ref{sec:toy_model}.

Our initial aim is to properly understand the evolution of the $P_{\delta_{x},\delta_{x}}^{L}$ term with respect to the global parameters of the EoR and on large physical scales; hence, we include the ``$L$'' superscript. We make use of standard perturbation theory (SPT) and consider the expansion of the ionisation fluctuation $\delta_{x} = \delta_{x}^{(1)}+\delta_{x}^{(2)}+ ...$, and pose the initial assumption that the linear fluctuation $\delta_{x}^{(1)}$ is initially small, Gaussian in nature, and follows the matter density fluctuation. This allows us to calculate the mode-coupling contribution of the $\delta_{x}$ field as we aim to capture its role in shaping the fluctuations of the field. 

We take note of the second-order ionisation fluctuation field term $\delta_{x}^{(2)}$, which can be re-expressed using second-order Eulerian perturbation theory as 
\begin{equation}
    \delta^{(2)}_{x} (\boldsymbol{k}) = \int \frac{d^{3} q}{(2 \pi)^{3}} \delta^{(1)}_{x}(\boldsymbol{q}) \delta_{x}^{(1)}(\boldsymbol{k}-\boldsymbol{q}) F_{2}(\boldsymbol{q_{1}},\boldsymbol{q_{2}}),
\end{equation}
where $F_{2}(\boldsymbol{q_{1}},\boldsymbol{q_{2}})$ is the second-order mode coupling kernel and can be expressed following \citet{Scoccimarro2000} as 
\begin{equation}
    F_{2}(\boldsymbol{q_{1}},\boldsymbol{q_{2}}) = \frac{5}{7} + \frac{\boldsymbol{q_{1}} \cdot \boldsymbol{q_{2}} }{2 q_{1}q_{2}} \Bigg(\frac{q_1}{q_2} + \frac{q_2}{q_1}\Bigg) + \frac{2}{7} \Bigg(\frac{\boldsymbol{q_{1}} \cdot \boldsymbol{q_{2}} }{q_{1}q_{2}}\Bigg)^{2}.
\end{equation} 
We utilise the perturbed field and solve for the autocorrelation power spectrum, noting that $\langle\delta_{x}^{(1)}, \delta_{x}^{(2)} \rangle$ vanishes for Gaussian initial conditions, as $P_{\delta_{x},\delta_{x}}^{L} \approx P_{\delta_{x}^{(1)},\delta_{x}^{(1)}}+P_{\delta_{x}^{(2)},\delta_{x}^{(2)}} + ...$ and so on. The first term $P_{\delta_{x}^{(1)},\delta_{x}^{(1)}}$ captures the linear growth of the field, and the second term, which captures the mode coupling of the field, can be expressed in the following form following Wick's theorem
\begin{equation}
\label{eq:p22}
    P_{\delta_{x}^{(2)},\delta_{x}^{(2)}} = 2 \int \frac{d^{3} q}{(2 \pi)^3}  (F_{2}(\boldsymbol{q},\boldsymbol{k}-\boldsymbol{q}))^{2}  P_{\delta_{x}^{(1)},\delta_{x}^{(1)}}(\boldsymbol{q})\, P_{\delta_{x}^{(1)},\delta_{x}^{(1)}} (\boldsymbol{k}-\boldsymbol{q}).
\end{equation}
We simplify Eq.~\eqref{eq:p22}, assuming isotropy and noting that for a linear perturbation, the variance of the field is $\sigma^{2}_{x} = \int \frac{d^{3} q}{(2 \pi)^{3}} P_{\delta_{x}^{(1)},\delta_{x}^{(1)}}$, to arrive at the following expression \citep[see sec.~3 of][for a similar derivation]{Lidz2007}
\begin{equation}
\label{eq::SPT_fluct}
P_{\delta_{x}^{(2)},\delta_{x}^{(2)}} \approx 2( \langle F_{2}^{2} \rangle) \sigma^{2}_{x} P_{\delta_{x}^{(1)},\delta_{x}^{(1)}}.
\end{equation}
The average value of the mode-coupling kernel $\langle F_{2} \rangle$ depends on the choice of approximation. Averaged over all angles, assuming statistical isotropy and $k \ll q$, it is $\langle F_{2} \rangle \approx \frac{17}{21}$. However, Eq.~\eqref{eq:p22} depends on the average of the squared kernel, $\langle F_{2}^{2} \rangle$, which is always larger than $\langle F_{2} \rangle^{2}$ and varies with $k$ and with the shape of the power spectrum. Moreover, effective field theory approaches show that once small-scale contributions are renormalised, the coupling between large and small scales is enhanced to $2 \times \frac{17}{21} = \frac{34}{21}$, scaled by the non-linear bias of the field \citep{McDonald2006, Assassi2014, Qin2026}. For strongly biased ionising sources, and with the higher-order terms neglected in our model, this coupling can be larger still. Since its exact value cannot be fixed within our minimal framework, we treat $\langle F_{2}^{2} \rangle$ as an effective coefficient $\omega$, which we calibrate to $\omega \in [2, 2.5]$ for $k \in [0.05, 0.5]$ cMpc$^{-1}$ and fix it to $\omega \approx 2$ for the remainder of this work\footnote{We have tested treating $\omega$ as a free parameter, marginalising over it with a wider uniform prior $\omega \in [1, 10]$, and found its effect on our constraints on the IGM parameters to be minimal.}.

To arrive at a generic approximation of the large-scale evolution of the ionisation field, we apply the following ansatz to the linear ionisation field on large scales such that
\begin{align}
\label{eq:bh}
\xHII^{(1)} &= \xa (1+b_g \delta),\mathrm{and}\, \delta_x = \xa b_g \delta,\\
P_{\delta_{x}^{(1)},\delta_{x}^{(1)}} &= \xa^{2} b_{g}^2 P_{\delta_{\rho},\delta_{\rho}},
\end{align}
where we assume the field traces the position of the ionising sources, which can be expressed in terms of the autocorrelation power spectrum of the density field and the source bias $b_g$, which we link to the halo bias, $b_g \approx b_h$. Following Eq.~\eqref{eq::SPT_fluct}, we arrive at the following form 
\begin{align}
\label{eq:pxx_lowk}
  P_{\xHII,\xHII}^{L}  &\approx  P_{\delta_{x}^{(1)},\delta_{x}^{(1)}}  + P_{\delta_{x}^{(2)},\delta_{x}^{(2)}}  \nonumber \\ &= \xa^{2} \big( 1+ 2 \omega \sigma^{2}_{x}\big) \,  b_{\mathrm{h}}^2 P_{\delta_{\rho},\delta_{\rho}}.
\end{align} 
We can interpret the role of the boosted variance term in Eq.~\eqref{eq:pxx_lowk} by comparing its evolution to that of the power spectrum of an ionisation field with randomly located bubbles that grow in number with the global ionisation fraction. In such a case, the power spectrum at $k \rightarrow 0$ cMpc$^{-1}$ evolves with the variance of the field $\sigma_{x}^{2} =\xa(1-\xa)$, following a parabolic function with a peak at $\xa = 0.5$. Conversely, the additional factor in Eq.~\eqref{eq:pxx_lowk} shifts the peak of the function for values of ionisation fraction to $\xa = \frac{1}{8} \Big(\sqrt{\frac{16}{\omega}+9}+3\Big) \approx 0.87-0.89$, matching the low-$k$ evolution seen in reionization models \citep[see fig. 3 of][]{Ghara2024}. 
\subsubsection{The transfer function of ionisation morphology and its effect on the scale dependence of the power spectrum of the ionisation field}
\label{sec:transfer_funct}
The ionisation field is defined by the creation, growth, and overlap of ionised regions during the EoR, which will critically determine the amplitude and scale dependence of the autocorrelation power spectrum across the EoR \citep{Xu2019}. To account for this in our parametrisation, we must quantify the role of the ionisation morphology and define an appropriate transfer function which couples the ionisation field to the underlying density field and captures the characteristic scale of ionised regions.

We define the survival probability of a packet of ionising photons travelling a distance $r$ as $S(r) = e^{-r/R}$, so that the term in the exponential defines the optical depth, so that $S$ is equal to unity at $r=0$ and $1/e$ at the characteristic scale $R$. To convert this expression to a probability density of absorption, we differentiate along the line of sight such that 
\begin{equation}
    p(r) = -\frac{dS(r)}{dr} = \frac{1}{R} \exp \Big(-\frac{r}{R}\Big).
\end{equation}
This probability integrates to unity and follows an exponential distribution of free paths, with a mean at the characteristic scale $R$. To convert this expression to describe a geometric distribution of shells, we divide by the volume element $dV = 4\pi r^{2} dr$ such that 
\begin{equation}
\label{eq:xi}
    \xi(r) = \frac{p(r)}{4\pi r^{2}} = \frac{1}{4\pi R } \frac{\exp(-\frac{r}{R})}{r^2}.
\end{equation}
The expression above is a useful descriptor of the 3D correlation function of ionised regions for several reasons. Firstly, it describes the distribution of absorption sites around an isotropic source in an absorbing medium, normalised to unity, and we adopt it as an effective kernel rather than the exact two-point function of the ionisation field. The units and the form itself are indicative of a number density of ionised regions, originating from the exponential distribution. Such a distribution factors in the overlapping of ionised regions and large abundance of small bubbles, which cannot be properly modelled when using a Gaussian or log-normal distribution alone \citep{Lin2016}. This contrasts with works such as by \citet{Hutter2021} where the cross-correlation function definition is a good description of the ionisation field up to ionisation fractions of around $\xa \approx0.20$ when overlap becomes important. Conversely, the exponential form adopted in this study expresses overlap through the free path distribution and is applicable during the EoR. Crucially, Eq.~\eqref{eq:xi} can also be better understood through the lens of percolation statistics discussed by \citet{Furlanetto2016}, where forms similar to $\xi(r)$ are shown to be descriptions of the number density of percolation clusters in 3D simulations. Moreover, similar expressions appear in phenomenological models such as in \citet{Mirocha2022}, where their eq.~30 is shown to converge to the same functional form.

We account for the fact that ionised regions trace the large-scale density field by taking the convolution of the ionisation kernel with the correlation function of the large-scale ionisation field, such that
\begin{equation}
\xi_{\xHII,\xHII}(r) = \left(\frac{1}{4 \pi R} \frac{e^{-r/R}}{r^2}\right) \ast \xi_{\xHII,\xHII}^{L}(r).
\end{equation}
We compute the spherically averaged power spectrum, noting that the convolution can be reduced to a multiplication in Fourier space
\begin{equation}
P_{\xHII,\xHII}(k) = 4\pi \int_0^\infty \xi_{\xHII,\xHII}(r) \frac{\sin(kr)}{kr} r^2 \, dr,
\end{equation}
\begin{align}
\label{eq::transfer_spectrum}
P_{\xHII,\xHII}(k) &= \frac{P_{\xHII,\xHII}^{L}}{kR} \int_0^\infty \frac{e^{-r/R}}{r} \sin(kr) \, dr \nonumber \\ &= \Bigg[\frac{\arctan(kR)}{kR} \Bigg] P_{\xHII,\xHII}^{L}. 
\end{align}
From Eq.~\eqref{eq:pxx_lowk}, we expand the power spectrum of the large-scale ionisation field, which is a function of the halo bias $b_{\mathrm{h}}$ and the matter density power spectrum $P_{\delta_{\rho},\delta_{\rho}}$, and includes the SPT correction from Sec.~\ref{eq:SPT} (setting $\omega \approx 2$), and we express it as
\begin{equation}
\label{eq:transfer_function}
     P_{x_{\mathrm{HII}},x_{\mathrm{HII}}}(k) = \Bigg[ \frac{\arctan(kR)}{kR}\Bigg]  \xa^{2} \big(1+ 2 \omega \sigma^{2}_{x}\big) \,  b_{\mathrm{h}}^2 P_{\delta_{\rho},\delta_{\rho}}. 
\end{equation}
For analytic convenience in the following calculations, we approximate the arctan function from Eq.~\eqref{eq::transfer_spectrum} with a form which, in the limit $kR \rightarrow 0$, is normalised to unity 
\begin{equation}
\label{eq:bub_model}
     P_{x_{\mathrm{HII}},x_{\mathrm{HII}}}^{\mathrm{ion}}(k) \approx \bar{x}_{\mathrm{HII}}^{2}  \frac{\frac{4}{\pi} (1+4\sigma^{2}_{x}) b_h^{2}}{\sqrt{\Big(\frac{4}{\pi}\Big)^{2}+(kR)^2}} P_{\delta_{\rho},\delta_{\rho}}. 
\end{equation}
This form closely matches its low-$k$ behaviour and recovers the $1/(kR)$ scaling at high $k$, and we add an ``ion'' superscript to denote that it is valid for ionised regions. Furthermore, Eq.~\eqref{eq:bub_model} takes a more recognisable form, similar to power-spectrum parametrisations in works such as \citet{Gorce2020} and \citet{Battaglia2013}. Moreover, taking the second derivative of the transfer function in Eq.~\eqref{eq:transfer_function} yields an inflexion point at $k \approx 1/ (1.214\,R)$, yielding a drop in power once the $k$-scale and bubble scale become comparable.
\subsubsection{The role of neutral islands in shaping the 21-cm field power spectrum}
\label{sec:toy_model}
The spherically averaged power spectrum measures fluctuations in the 21-cm field during the EoR. In most analytic EoR models, these fluctuations are attributed solely to the formation and growth of ionised bubbles \citep[though see][]{xu2014analytical}. However, the hydrogen field that governs the 21-cm signal is a two-phase medium, consisting of ionised and neutral hydrogen. In this section, we investigate how the intensity of the fluctuations of the neutral IGM will impact the 21-cm power spectrum. 

\begin{figure}
    \centering
    \includegraphics[width=\columnwidth]{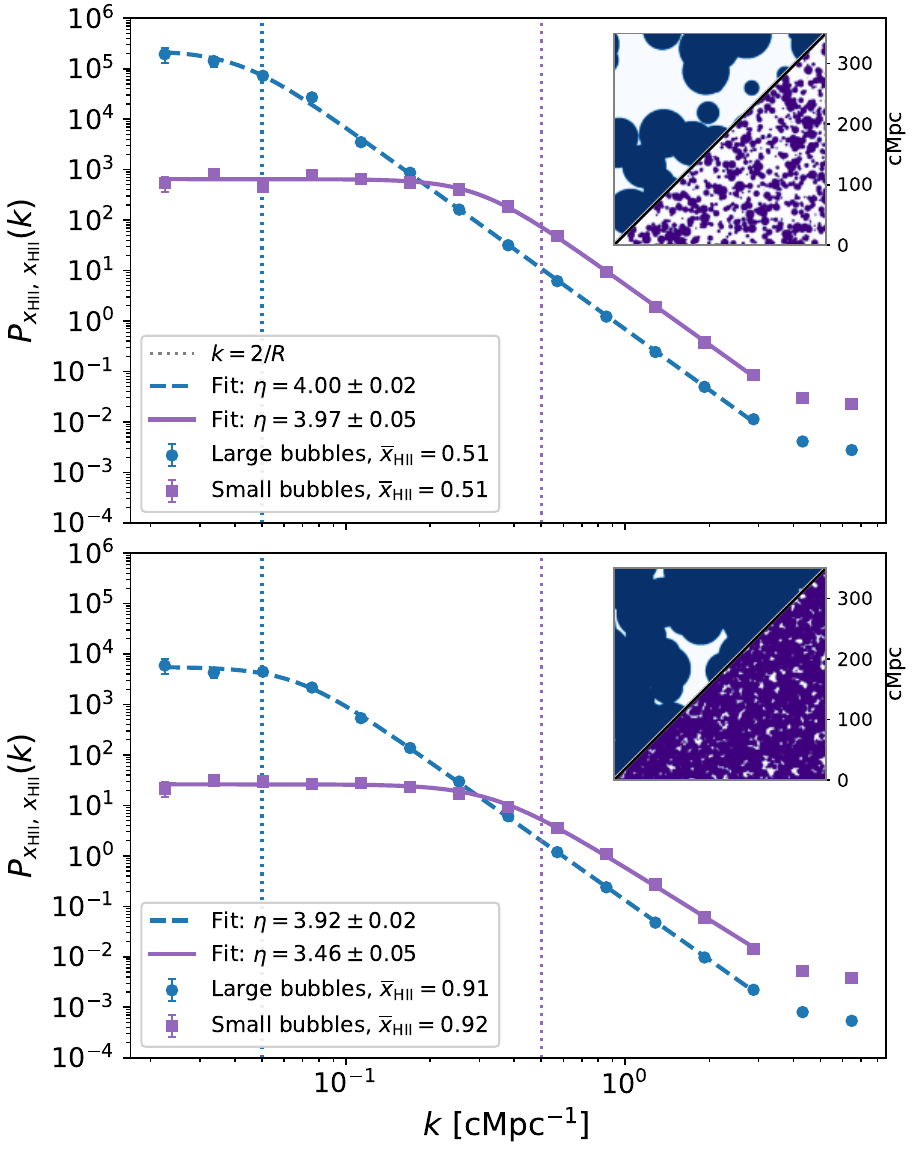}
    \caption{The top right of each panel shows two toy-model ionisation fields, in which bubbles of a fixed mean size are randomly distributed across a 3D grid. The bubble sizes in the two models are $4$ cMpc and $40$ cMpc, hence the small- and large-bubble nomenclature. Each panel shows the autocorrelation power spectra of the ionisation field for the large and small bubble models, shown as circles and squares, respectively. The top panel is set at the midpoint of reionization ($\xa \approx 0.51$) and the bottom panel is towards its final phases ($\xa \approx 0.91$). Each power spectrum is fitted with a simple power-law model $A / (1+(kR_b)^{\eta})$, shown as a dashed/solid line, where $\eta$ is the spectral slope.}
    \label{fig:PS_neutral_Islands_casestudy}
\end{figure}

To understand the role of the neutral islands on the 21-cm power spectrum, we construct the following example. We focus primarily on the ionised/neutral field and create a toy model in which we randomly populate ionised bubbles on a 3D grid, following a Gaussian distribution, to a predetermined ionisation fraction. We focus on two models, a large bubble model where the mean bubble size is $R_b = 40$ cMpc and a small bubble model where $R_b = 4$ cMpc. Both distributions are given a small spread around the mean of around $2$ cMpc. In Fig.~\ref{fig:PS_neutral_Islands_casestudy} we highlight two examples of autocorrelation power spectra of the ionisation field as well as their corresponding maps, in the top panel at the midpoint and another at the very final stages of reionization. The former is chosen to mark the epoch when both neutral and ionised fluctuations in the power spectrum are comparable, while the latter is chosen to highlight the role of neutral islands.
On the left-hand side of both panels, we present the autocorrelation power spectra of the ionisation/neutral fraction field. We perform a simple $\chi^2$ minimisation on each power spectrum, weighting each $k$-bin by its sample variance, $\sigma_P = P\sqrt{2/N_k}$, where $N_k$ is the number of modes in the bin. We fit the power spectra with a function of the form $A / (1+(kR_b)^{\eta})$ and focus on the spectral slope parameter ($\eta$). For a simplistic model such as this, the power-spectrum slope should be $\eta = 4$ for non-overlapping bubbles \citep[see sec.1 of][for a derivation]{Gorce2020}.
The ionisation autocorrelation power spectra at $\xa \approx 0.5$ shown in the upper left panel of Fig.~\ref{fig:PS_neutral_Islands_casestudy} confirm the status quo and our initial intuition. Both models show slope values close to the expected value, and recover ionised radii close to the input distribution. The small scatter can be attributed to both the accuracy of the $\chi^{2}$ minimisation and the role of overlapping structures. This intuition remains largely the same when considering the shape of the large bubble model's power spectrum at $\xa \approx 0.92$ in the bottom panel of Fig.~\ref{fig:PS_neutral_Islands_casestudy}. The slope is slightly less steep at $\eta = 3.92 \pm 0.02$. Nevertheless, this can be attributed to both the high degree of overlap and high variance at low-$k$ values. However, a notable change has occurred when viewing the slope of the small bubble model, which is now around $\eta = 3.46 \pm 0.05$, flatter than expected. We can understand the additional power in the following manner. In the final phases of reionization, when $\xa> 0.85$, we observe a de-percolation of the neutral field \citep{Furlanetto2016,Giri2019}. In this period, the large neutral cluster surrounding the ionised bubbles is broken into neutral islands. Since the autocorrelation power spectrum measures the positive contribution of the fluctuation of the field, the expression of $\langle \delta_{\HII},\delta_{\HII} \rangle$ will be positive both for ionised bubbles where $\delta_{\HII} > 0$ and neutral islands where $\delta_{\HII} = -1$. In turn, the small, numerous neutral islands in this model have a size of approximately a megaparsec, smaller than that of the ionised bubbles, and will contribute additional power at scales of $\approx2/R_n$, thus boosting the power and flattening the spectral slope. Conversely, this effect is less pronounced for the large bubble model. At a fixed global ionisation fraction, the same volume of neutral hydrogen is now centred in fewer, larger neutral islands, whose size and frequency of occurrence have a much weaker effect on the ionisation power spectrum than the ionised regions. 

Therefore, we can quantify the role of neutral islands on the ionisation field autocorrelation power spectrum, following by analogy in Sec.~\ref{eq:SPT} \& \ref{sec:transfer_funct} to arrive at the expression
\begin{equation}
\label{eq:HI_model}   P_{x_{\mathrm{HI}},x_{\mathrm{HI}}}^{\mathrm{neut}}(k)   \approx \bar{x}_{\mathrm{HI}}^{2}  \frac{ (1+4\sigma^{2}_{x}) b_v^{2}}{1 +(kR_n)^4} P_{\delta_{\rho},\delta_{\rho}}. 
\end{equation}
The above equation can be derived following the same logical steps performed to arrive at Eq.~\eqref{eq:bub_model}, except for three differences. Firstly, the fluctuation must be multiplied by the square of the global neutral fraction, to ensure that the autocorrelation power spectrum is normalised at the start and end of reionization. Naturally, when $\bar{x}_{\mathrm{HI}}^{2}  \rightarrow 0$ the spectrum from Eq.~\eqref{eq:HI_model} also diminishes. Conversely, early on, when the global neutral fraction is close to unity, the effective size of neutral islands $R_n \rightarrow \infty$, and the power spectrum approaches zero. 
Secondly, the transfer function in Eq.~\eqref{eq:bub_model} describes a percolated field, whose ionised regions extend beyond individual \HII bubbles. Conversely, the neutral morphology is driven by fewer ``non-overlapping'' and isolated \HI regions. Therefore, the transfer function most suitable for describing the neutral islands has the spectral shape $1 / (1+(kR_n)^{4})$, as seen in Fig.~\ref{fig:PS_neutral_Islands_casestudy}, since we are describing $\Hi$ islands after the de-percolation of the neutral cluster. Despite this, Eq.~\eqref{eq:HI_model} remains convolved with the density power spectrum because, albeit relatively disconnected, the large-scale diffuse distribution and morphology of the neutral islands trace the large voids, which themselves trace the density field itself\footnote{Only in the final fraction of a per cent of neutral hydrogen, when the number of neutral islands becomes sufficiently small, does their large-scale distribution resemble Poisson noise.}. Lastly, although we include the role of the void bias $b_v$ in the above expression, we effectively choose $b_v \approx 1$ during the final phase of reionization. We choose this simplification since at redshifts as high as six and above, very large voids are mostly rare \citep{DAloisio2007}, and likely the large-scale void bias is shot-noise dominated \citep{Hamaus2014}.
\subsubsection{The assembled IGM parametrisation}
\label{sec::IGM-params}
We can reform every term in Eq.~\eqref{eq:Decomposition}, combining Eq.~\eqref{eq:bub_model} \& \eqref{eq:HI_model} to the following form
\begin{equation}
\label{eq:p21_model}
    P_{21{\rm cm}} = T_{0}^{2} \bar{x}_{\mathrm{HI}}^{2} \Bigg( \frac{\big(\frac{4}{\pi}\big)(1+b_{i}(f+1))^{2}}{\sqrt{\big(\frac{4}{\pi}\big)^{2}+(kR_{i})^{2}}} + \frac{1+4\sigma_{\mathrm{xHII}}^{2}}{{1+(kR_{n})^{4}}} \Bigg) P_{\delta_{\rho},\delta_{\rho}}, 
\end{equation}
where the additional term $b_i = - \frac{\bar{x}_{\mathrm{HII}}}{\bar{x}_{\mathrm{HI}}} \sqrt{1+4\sigma_{\mathrm{xHII}}^{2}} b_h$ includes the factor to transform the ionisation fluctuation from Eq.~\eqref{eq:bub_model} in terms of the neutral field fluctuation, whilst $f$ accounts for the contributions of the cross terms in Eq.~\eqref{eq:Decomposition} (see Appendix~\ref{sec:cross-terms} for a derivation and discussion). Therefore, Eq.~\eqref{eq:p21_model} follows a two-phase medium, describing the growth of large percolated ionised regions and the presence of neutral islands. Note that the two terms are not independent fields. However, because each term describes a different morphological component, they dominate at opposite ends of the EoR, so jointly fitting the parameters selects the region of parameter space where their dependence is consistent with the data. Furthermore, we apply the $f$ term only to the ionised component. This choice is motivated by the fact that the ionisation morphology is more strongly biased by the density field, since reionization is driven by sources at the highest-density peaks. The neutral islands instead occupy the voids, where we take $b_v\approx1$ following Sec.~\ref{sec:toy_model}, and their coupling to the density field is weaker. The neutral term becomes relevant only toward the end of the EoR, where the cross terms are subdominant (see Appendix~\ref{sec:cross-terms}). Lastly, the bracketed term of the ionisation component $(1+b_i(f+1))^2$ passes through zero at a given ionisation fraction, where the growing ionisation fluctuation exactly cancels the density fluctuation it traces. Physically, this is the epoch of equilibration, when the sizes of ionised regions have grown to mask out the highest density peaks, suppressing the large-scale 21-cm amplitude \citep{Lidz2008}.
\subsection{Redshift-dependent model of the 21-cm power spectrum}
\label{sec:IGM_params}
In Section~\ref{sec:21ps_parametrisation}, we quantified how the IGM parameters impact the amplitude and shape of the 21-cm power spectrum during the EoR. While our analysis was shown at a fixed redshift (and therefore a fixed neutral/ionisation fraction), the processes that shape the IGM parameters are both scale-dependent and redshift-dependent. Understanding how the IGM parameters evolve will therefore not only allow us to model the 21-cm power spectrum across different stages but also provide further insight into reionization and break degeneracies in the IGM parameter space. In this section, we parametrise the redshift evolution of the IGM quantities. We refer to these parametrisations as the ``evolution parameters'' to distinguish them from the IGM parameters. We aim for simple parametrisations already used in the literature, with physical interpretations, to avoid overfitting and allow clear comparison with the data. Ultimately, we can marginalise over these parameters.
\subsubsection{Volume and mass-weighted reionization history}
\label{sec:ion_history}
A crucial parameter shaping the evolution of the 21-cm signal is the ionisation fraction $\xa$. This is most clearly seen in Eq.~\eqref{eq:p21_model}, however, the role of the ionisation fraction is also fundamental in shaping other IGM parameters such as the sizes of ionised and neutral regions as well as the mass-weighted ionisation fraction ($\xc$).

Accurate and precise models of the ionisation history are essential in estimating the Thomson optical depth in observations of the CMB \citep[see][]{Planck2020}. We settle on a modified version of the asymmetric parametrisation described by \citet{Douspis2015} and \citet{Georgiev2024}
\begin{align}
    \label{eq:reion_parametrisation}
    z &< z_{\mathrm{re}} - dz, \quad \bar{x}_{\mathrm{HII}}(z) = 1,\\
    z_{\mathrm{re}} - dz &\leqslant z < z_{\mathrm{early}}, \quad \bar{x}_{\mathrm{HII}}(z) = \frac{1}{2} + \frac{1}{2}\,\frac{t(z)^{\alpha_{\mathrm{EoR}}} - t(z_{\mathrm{re}})^{\alpha_{\mathrm{EoR}}}}{1 - t(z_{\mathrm{re}})^{\alpha_{\mathrm{EoR}}}},\nonumber\\
    z &\geqslant z_{\mathrm{early}}, \quad \bar{x}_{\mathrm{HII}}(z) = 0,\mathrm{\,where\,}  t(z) = \frac{z_{\mathrm{early}} - z}{z_{\mathrm{early}} - z_{\mathrm{re}} + dz}, \nonumber
\end{align}
where $\alpha_{\mathrm{EoR}}$ is the evolution/rate of the ionisation history and $z_{\mathrm{early}} = 20$ is fixed to a period where $\bar{x}_{\mathrm{HII}} \approx 10^{-4}$. Such a model is naturally asymmetric, modulated by $\alpha_{\mathrm{EoR}}$. The remaining two free parameters are the midpoint of reionization $z_{\mathrm{re}}$, normalised by construction at $\bar{x}_{\mathrm{HII}} =0.5$, and the duration of reionization $dz$. Constraining $z_{\mathrm{re}}$ is quite advantageous as the midpoint coincides with the redshift at which the 21-cm power spectrum is at its highest amplitude \citep{Georgiev2022}. We choose such a parametrisation as its inferred values are naturally complementary to CMB observables such as the Thomson optical depth \citep{Planck2020} and the patchy kinetic Sunyaev-Zel'dovich effect \citep{Reichardt2021,Chaubal2026}. Lastly, these parameters also allow us to define and thus constrain the end of reionization $z_{\mathrm{end}} = z_{\mathrm{re}}-dz$, which is reasonably constrained by the \lya forest observations of the final phases of reionization with QSO spectra \citep{Bosman2022} and can also separately be inferred from the amplitude of the 21-cm power spectrum \citep{giri2024end}. Therefore, the aforementioned probes can naturally be folded into our analysis to provide more stringent constraints on the IGM parameters. Note that this parametrisation is clipped such that $\xa \ge 0$, which affects redshifts above the range of this study.

Another parameter essential for understanding the role of the IGM parameters is the mass-weighted ionisation fraction $\xc = \overline{\rho x_{\mathrm{HII}} }/\overline{\rho}$. Moreover, as discussed in Appendix~\ref{sec:cross-terms}, the mass-weighted ionisation fraction plays an essential role in expressing the higher-order terms in Eq.~\eqref{eq:Decomposition} \citep[see sec.~3.2 of][]{Lidz2007}. We express the mass-weighted ionisation fraction in the following form
\begin{equation}
\label{eq:xhii_m}
    \xc = \xa^{\gamma},\mathrm{\,where\,\,}\gamma>0.
\end{equation}
The simple power law is advantageous because it describes a large-scale self-similar process, whose growth or suppression is regulated by the factor $\gamma$. Hence, such a factor is informative about the large-scale correlation between the matter density and ionisation field and how ``inside-out'' or ``outside-in'' the process of reionization is \citep[see][for a discussion]{Iliev2014}. Similar functional forms have also been found by \citet[][see sec.~3.2]{Oku2025} from the study of both volume- and mass-weighted fractions for hydrodynamic simulations of the EoR. The parameter range of $\gamma$ can be further constrained. Given that $ \xc = \overline{x_{\mathrm{HII}} \rho}/\bar{\rho}$, it follows that $\gamma = 1$ corresponds to the case where density and ionisation fields are uncorrelated and hence unphysical. Moreover, the set of simulations used within this study is based on the assumption that the EoR is an inside-out process, i.e. it is the denser peaks that reionised first while the lowest-density regions reionised later in time. A toy model example was used to create a mock inside-out simulation, redistributing the global ionisation fraction of this model to a ranked list of the density field. For the case of an extremely inside-out reionization we find $\gamma \approx 0.5$, whilst for an extreme outside-in model the value is closer to $\gamma \approx 1.5$. Randomly assigning ionised cells to the density field yields the expected $\gamma \approx 1$. This form is also complementary to 21-cm-LAE cross-correlation studies \citep{Pagano2021} whilst constraints on $\xc$ are complementary to observations of the global 21-cm signal from EDGES and SARAS \citep{Monsalve2017,Singh2018}.
\subsubsection{Clustering of sources}
\label{sec:clustering}
The clustering of the first ionising sources is a key driver of the reionization morphology \citep{Furlanetto2004,McQuinn2018}, and enters our IGM framework through the halo bias in Eq.~\eqref{eq:bh}. We include it among the IGM parameters because it affects our model only through the large-scale amplitude of the ionisation field in Eq.~\eqref{eq:pxx_lowk}, via how strongly the ionisation morphology correlates with the underlying density field. We therefore constrain the imprint of the sources on the IGM, rather than the sources themselves. This clustering is set by that of the dark matter (DM) halos hosting the sources, and hence by the underlying cosmology.

The clustering of these ionising sources is best understood by modelling the halo (and hence source) bias discussed in Sec.~\ref{sec:21ps_parametrisation}. In the IGM framework, we generate $b_h$ with the {\sc COLOSSUS} toolkit \citep{Diemer2018}, using the bias and mass function for the cosmology of each simulation and taking the mass-weighted bias of all halos above $M_{\mathrm{min}}$. For visual purposes, we can analytically showcase the linear halo bias form following \citet{Cole1989,Mo1996,Sheth2001}
\begin{align}
\label{eq:halo_bias}
b_h(M,z)&=\!\left[1+\frac{a\nu^2-1}{\delta_c}+\frac{2p}{\delta_c\!\left(1+(a\nu^2)^p\right)}\right], \nonumber \\ 
\nu(M,z) & = \frac{\delta_c}{\sigma(M)\,D(z)},
\end{align}
and $\delta_c=1.686$ is the critical overdensity in linear perturbation theory, $D(z)$ is the linear growth factor and $\sigma(M)$ the root mean square fluctuation of the density field on the scale of a halo of mass $M$, where $a=0.707$ and $p=0.3$ are taken from the Sheth–Tormen halo mass function \citep{Sheth1999}, which is known to provide a poor fit to simulated mass functions at high redshift \citep{Watson2013}. In this simplified form, $b_h$ is a function of two parameters, the redshift and the mass $M$, which is the mass of the dark matter halos considered. Hence, not only can $b_h$ provide the redshift dependence of this IGM parameter, but we can go beyond its emergent nature and link it to a property of reionization such as the minimum mass of halos capable of hosting ionising sources ($M_{\mathrm{min}}$). 

There exist several caveats when discussing halo bias and its role in shaping the 21-cm power spectrum. Firstly, simulations of reionization model the density field differently, impacting the amplitude and evolution of the halo bias, while the linear bias from {\sc COLOSSUS} neglects the non-linear source clustering captured by simulations. The \citet{Sheth2001} bias, fitted to $N$-body models, differs slightly from estimates such as \citet{Tinker2010}. Moreover, astrophysical feedback modulates the clustering by determining which halos host ionising sources. Models of reionization exist where the sources are more stochastic \citep{Nasirudin2020} or where $M_{\mathrm{min}}$ changes over time \citep[see fig.5 of][]{Katz2019}. We elaborate on this in Sec.~\ref{sec:polar_results}. 
\subsubsection{Parametrisations of the growth of neutral and ionised regions}
\label{sec:R_params}
Defining the growth of ionised regions during the EoR is a non-trivial task, yet the role of percolation of ionised structures plays an important role, as it drives an exponential-like growth \citep{Stauffer2020} observed in analytical excursion-set based approaches such as that by \citet{Furlanetto2005} and in bubble sizes from hydrodynamic simulations \citep{Neyer2024,Jamieson2025}. Hints of this behaviour are also noted in bubble size estimates from high-redshift galaxy observations \citep{harikane2023comprehensive, Hayes2023, Umeda2023}.

Therefore, we choose a model where the size of ionised regions follows the shape
\begin{equation}
\label{eq:Ri}
    R_i=10^{\alpha_i}\,\exp\!\Bigg[-\beta_i\Bigg(\frac{1-\xa}{1-x_0} - 1\Bigg)\Bigg]\, \mathrm{cMpc,\,where}\, x_0 = 0.5.
\end{equation}
The model is tuned for a process of exponential growth and designed to be a function of the ionisation fraction rather than redshift. The $x_0$ parameter acts as a pivot for the parametrisation, shifting the constraining power of the model to the midpoint of reionization. The reason for pivoting the model at the midpoint is twofold. Firstly, the characteristic feature that $R_i$ affects the 21-cm power spectrum is around $k \approx 0.1$ cMpc$^{-1}$, among the scales where current interferometers place their tightest constraints. Secondly, the midpoint is a period where the cross-terms contribute least to the total power, as the one-point distribution of the 21-cm brightness temperature field is closer to Gaussian \citep{Mellema2006b,Georgiev2022}, so the feature is more cleanly attributable to the ionisation morphology. The fit is designed such that the ${\alpha_i}$ parameter constrains $R_{i}$ at the midpoint of reionization, to compare with the aforementioned observations, a key redshift range for telescopes such as JWST. The ${\beta_i}$ parameter governs the growth of $R_{i}$ throughout reionization and should capture the suppression due to the scale of ionised regions imposed by the mean free path parameter used in simulations, while values of $\beta_{i}$ lower than unity in Eq.~\eqref{eq:Ri} allow for an evolution more similar to a power law, yielding some level of flexibility. Most importantly, it is a simple metric for comparison with other, more physically motivated models. 

Following a similar philosophy, we define the neutral patch size as
\begin{equation}
\label{eq:Rn}
R_n = 10^{\alpha_n}\Bigg(\frac{\xa}{x_1}\Bigg)^{-\beta_n}\, \mathrm{cMpc}, \quad \mathrm{where}\ x_1 = 0.85.
\end{equation}
The motivation is similar, with a few caveats. First, the shape differs to reflect the gradual ionisation of neutral islands. The differences also serve to distinguish the models in the fitting. The model is pivoted towards the end of reionization ($\xa \approx 0.85$), when the neutral field de-percolates and where the size and shape of detected neutral islands \citep{Becker2015} should contain information on the EoR process \citep{Giri2019}.

We summarise the evolution parameters in Table~\ref{tab:priors}, and in Sec.~\ref{sec:mcmc} we discuss our prior ranges. Note that the IGM framework is modular by design, and the functional form chosen for any parameter can be exchanged or improved independently of the rest. For example, the EoR history in Eq.~\eqref{eq:reion_parametrisation} could easily be replaced by a Weibull function \citep{Trac2022}.
\section{Simulations}
\label{sec:sims}
We test the applicability of our IGM-based framework against models of the 21-cm power spectrum generated from simulations of the EoR. We outline three simulation codes below that adopt different philosophies for generating the 21-cm field, which will help us test the robustness of our predictions and highlight the framework's assumptions.
\subsection{\texorpdfstring{C$^{2}$-Ray: 3D Radiative Transfer + $N$-body density field}{C2-Ray: 3D Radiative Transfer + N-body density field}}\label{sec:C2-Ray}
The halo and density fields are generated with the $N$-body simulation {\sc CUBEP$^{3}$M} code \citep{Harnois2013} and post-processed using the 3D radiative transfer code {\sc $\mathrm{C}^2$-Ray} \citep{Mellema2006a,iliev2006}. The halos in the simulation are assigned an ionising luminosity proportional to their mass and to the halo growth rate. The source model is designed such that the total number of ionising photons emitted up to a given redshift is directly proportional to the fraction of matter collapsed into dark matter halos at that redshift \citep[see][]{Schwandt2026M}. 

The {\sc C$^{2}$-Ray} simulation used within this work was first introduced in sec.~3.1 of \citet{giri2021measuring} as \texttt{FN1}, which we will label \texttt{C2R}. The simulation has a comoving volume with a box length of $L_{\mathrm{box}} = 500 h^{-1}$~cMpc$\approx 714$~cMpc, using $N_{\mathrm{particles}} = 6912^{3}$ with a particle mass of $M_{\mathrm{particles}} = 4.05 \times 10^{7} M_{\odot}$. The halos hosting ionising sources were identified on the fly through the spherical over-density halo-finder method \citep{Watson2013}, resulting in a minimum halo mass of about $10^9 M_{\odot}$, i.e. only high-mass atomically cooling halos. The model has a constant ionising efficiency $\zeta = 50$ \citep[see tab.~1 of ][]{giri2021measuring}, which is the number of ionising photons escaping into the IGM per baryon collapsed in these halos\footnote{This simulation was run as part of the PRACE (Partnership for Advanced Computing in Europe) project {\sc PRACE4LOFAR}.}. The radiative transfer is processed by \citet{Mellema2006a} on a $250^{3}$ grid. The cosmology parameters ($\Omega_{\mathrm{m}},\Omega_{\mathrm{b}},h,n_{s},\sigma_{8}$) = (0.27, 0.044, 0.7, 0.96, 0.8) are consistent with WMAP \citep{hinshaw13} and Planck \citep{planck14}.
\subsection{\texorpdfstring{{\sc 21cmFAST}: Semi-Numerical Radiative Transfer + 2LPT density field}{{\sc 21cmFAST}: Semi-Numerical Radiative Transfer + 2LPT density field}}\label{sec:21cmFAST}
We use version 4.3.2 of the {\sc 21cmFAST} code \citep{Mesinger2011, Davies2025}, which generates the matter density field with second-order Lagrangian perturbation theory (2LPT) while the process of reionization is simulated utilising the excursion set based approach first introduced in \citet{Furlanetto2004}. We run a simulation with an identical box size and cell size to that of the simulation in Sec.~\ref{sec:C2-Ray} to sample a large box at matching resolution. However, we keep the default cosmology of the code \citep{Planck2020}, such that $(\Omega_{\rm m}, \Omega_{\rm b}, h, n_s, \sigma_8) = (0.310, 0.0490, 0.677, 0.967, 0.810)$, and do not include inhomogeneous recombinations. Ionising sources are hosted by halos above a sharp cut-off of $M_{\mathrm{min}} = 10^{10} M_{\odot}$, notably higher than in the \texttt{C2R} case. We do this to create variance in the IGM properties so as to further stress-test our IGM parametrisation. Crucial to this test is the different source parametrisation, for which we use a constant ionising efficiency $\zeta = 100$. Note that we set $R_{\mathrm{max}} = 100$ cMpc, pushing the features it imprints on the power spectrum to $k <2/R_{\mathrm{max}} = 2\times 10^{-2}$ cMpc$^{-1}$.
\subsection{\texorpdfstring{{\sc POLAR}: 1D Radiative Transfer + $N$-body density field}{POLAR: 1D Radiative Transfer + N-body density field}}\label{sec:Polar}
The {\sc POLAR} code \citep[see][for a detailed description]{Ma2023,Ma2026} utilises the {\sc Jiutian-300} $N$-body dark matter simulation \citep{Han2025}, run with {\sc GADGET-4} \citep{Springel2021} using the cosmological constraints consistent with \citet{Planck2020}, where $(\Omega_{\rm m}, \Omega_{\rm b}, h, n_s, \sigma_8) = (0.311, 0.049, 0.6766, 0.9665, 0.802)$. The DM simulation has a box length of approximately 443 cMpc, using $N_{\mathrm{particles}} = 6144^{3}$ with a particle mass of $M_{\mathrm{particles}} = 1.5\times 10^{7} M_{\odot}$. The dark matter halos and sub-halos are identified following \citet{Springel2001,Springel2005}, retrieving a minimal resolved halo mass around $10^{8} M_{\odot}$. Rather than a parametrisation linking the number of ionising photons to the mass of halos, such as in the aforementioned methods, {\sc POLAR} uses ionising sources assigned from the semi-analytic galaxy formation model {\sc L-GALAXIES} \citep{Henriques2020}. This captures a wide range of physical and galactic properties, tracing galaxy mergers through the constructed merger trees and including mechanisms that impact star formation, such as AGN feedback and tidal effects \citep{Ma2025}. The radiative transfer is processed by the 1D RT Grizzly code \citep{Ghara2015a} on a $256^{3}$ grid, which takes into account galaxy properties such as the age of the galaxy and its stellar mass. The 1D ionisation profiles are calculated for each galaxy, while the ionisation field is constructed through fast Fourier transforms. The simulation used within this study is the fiducial model from \citet{Ma2026} (see tab.~1), which we will refer to as \texttt{PLR}.
\section{MCMC-sampling}
\label{sec:mcmc}
We apply the 21-cm power spectrum model from Eq.~\eqref{eq:p21_model}, where we sample the IGM parameters through their evolution parameters discussed in Sec.~\ref{sec:IGM_params} and study the ratio between the 21-cm and matter density power spectra from the simulations in our analysis. 
\begin{table}
    \centering
    \begin{tabular}{lcc}
        Parameter & Label & Prior \\
        \hline\hline
        \textbf{Ionisation history} & $\xa$ & \\
        \cline{1-2}
        EoR midpoint & $z_\mathrm{re}$ & [4.5, 15.0] \\
        EoR duration & $dz$ & [0.01, 4.1] \\
        EoR evolution rate & $\alpha_{\mathrm{EoR}}$ & [0.01, 30.0] \\
        \noalign{\smallskip}
        \textbf{Mass-weighted ionisation history} & $\xc$ & \\
        \cline{1-2}
        Density-ionisation correlation & $\gamma$ & [0.0, 1.0] \\
        \noalign{\smallskip}
        \textbf{Clustering of sources} & $b_h$ & \\
        \cline{1-2}
        Minimum mass of halos hosting sources & $\log_{10}(M_{\mathrm{min}})$ & [7.0, 12.0] \\
        \noalign{\smallskip}
        \textbf{Ionised regions} & $R_i$ & \\
        \cline{1-2}
        $\HII$ bubble radii at $\xa \approx 0.5$ & $\alpha_{i}$ & [0.1, 2.4] \\
        $\HII$ bubble radii evolution rate & $\beta_{i}$ & [0.01, 4.5] \\
        \noalign{\smallskip}
        \textbf{Neutral regions} & $R_n$ & \\
        \cline{1-2}
        $\HI$ patch radii at $\xa \approx 0.85$ & $\alpha_{n}$ & [0.1, 2.4] \\
        $\HI$ patches radii evolution rate & $\beta_{n}$ & [0.1, 3.5] \\
        \hline 
        \hline 
    \end{tabular}
    \caption{Evolution parameters of the IGM framework and their imposed flat prior ranges, grouped by the IGM parameter each set describes.}
    \label{tab:priors}
\end{table}
We present the flat prior bounds in our IGM framework in Table~\ref{tab:priors}. A strength and benefit of this framework comes from the fact that its parameters are physically significant and thus the prior ranges can be constrained. Nevertheless, the prior placed on the midpoint of reionization $z_{\mathrm{re}}$ is chosen here to broadly encapsulate the full EoR regime. Similarly, the duration of reionization is only limited in its upper bound of $dz < 4.1$ from estimates from the patchy kinetic Sunyaev-Zel'dovich measurement by \citet{Reichardt2021,Chaubal2026}. Note, however, that $dz$ in Eq.~\eqref{eq:reion_parametrisation} spans a period from the $z_{\mathrm{re}}$ to the end of the EoR. Conversely, the pkSZ measurements refer to the interval between 25$\%$ and 75$\%$ ionisation. Since our recovered histories re-ionise rapidly during the final stages, the $dz$ parameter we define spans a smaller redshift interval than the 25-75$\%$ definition in the pkSZ measurements, making this bound conservative. The reionization rate parameter $\alpha_{\mathrm{EoR}}$ is afforded a large prior range, excluding only values where $\alpha_{\mathrm{EoR}}\leq0$, which are unphysical in their definition in Eq.~\eqref{eq:reion_parametrisation}. We choose to limit the density-ionisation parameter by $0 < \gamma \leq 1$, as a null value is unphysical on inspection of Eq.~\eqref{eq:xhii_m} while a value of unity is indicative of an uncorrelated/random reionization scenario (i.e. $\xa =\xc$). The limits on the minimum mass of the halos producing ionising photons are chosen to be close to the atomically cooling threshold \citep[e.g.][]{Nebrin2023}. Moreover, the parameters defining the size of the characteristic ionised bubble at the midpoint of the EoR in Eq.~\eqref{eq:Ri} cannot be null and must be below post-EoR measurements of the mean free path of ionising photons \citep{Worseck2014, Zhu2023}. Our constraints on the size at that period allow for bubble sizes of values from unity to $250$ cMpc, well above expected values inferred from galaxy observations \citep{Hayes2023} yet consistent with large lines-of-sight \citep{Helton2026,Yue2026}, which in themselves can be applied to further constrain the prior. Conversely, the model for the sizes of neutral islands in Eq.~\eqref{eq:Rn} is defined toward the end of the EoR \citep{Becker2015,Keating2020K} to include neutral islands with radii up to $250$ cMpc.

The true values of the evolution parameters of the ionisation history are either measured directly from the simulation output (e.g. $z_{\mathrm{re}}$), or inferred by fitting Eq.~\eqref{eq:reion_parametrisation} for $\xa$ and Eq.~\eqref{eq:xhii_m} for $\xc$. Similarly, the halo bias is obtained either directly from the simulation set-up (such as the stated $M_{\mathrm{min}}$) or from the ratio of the halo and matter autocorrelation power spectra. The sizes of $\Hi$ and $\HII$ regions cannot be readily extracted from the simulation output. For ionised regions, we use the radial mean free path bubble size distribution (RMFP-BSD) method from Georgiev et al. (2026b, submitted), which casts six rays along the Cartesian axes from a random origin to a neutral hydrogen wall, yielding a metric $R_i$ with a noticeably tighter spread around its median. For neutral regions we use the MFP-BSD method, since they are naturally more asymmetric. For completeness, we also include $R_i$ at $\xa>0.85$ and $R_n$ at $\xa<0.50$, although an ionised region is hard to define once neutral regions de-percolate into \HI islands, as is a neutral region in the early EoR. Finally, the characteristic peak in the 21-cm power spectrum and that in the MFP-BSD need not measure the same morphological aspect, and must be interpreted with care.

We construct a Markov Chain Monte Carlo (MCMC) sampler using the {\sc emcee} Python package \citep{Foreman-Mackey2013}. We evaluate the validity of our model by assuming a Gaussian likelihood with a fixed uncorrelated error across all of our data
\begin{equation}
    \ln{\mathcal{L}_{s}(\theta)} = -\frac{1}{2}\sum_{k} \Bigg[ \Bigg(\frac{1}{\sigma}\log_{10}\frac{P^{\mathrm{data}}(k)}{P^{\mathrm{model}}(k)}\Bigg)^{2} + \ln{2\pi\sigma^{2}} \Bigg]
\end{equation}
where $P \equiv |P_{21}/(T_{0}^{2}P_{\delta\delta})|$ and $\sigma$ is the logarithmic uncertainty across our data points and is currently fixed at $\sigma = 0.2-0.3$ dex, primarily motivated by the theoretical uncertainty inherent in simplifying the cross terms in Appendix~\ref{sec:cross-terms}. As this uncertainty is multiplicative in origin, it is naturally symmetric in log-space. Such a fixed treatment is purposefully conservative for a theoretical work such as this, leaving a Gaussian likelihood in $P(k)$ with the appropriate thermal noise covariance for future studies of real observations.

Furthermore, we allow each redshift to be weighted in the likelihood, such that
\begin{equation}
    \ln{\mathcal{L}_{\mathrm{tot}}(\theta)} = \Sigma_{s} w_{s} \ln{\mathcal{L}_{s}(\theta)}
\end{equation}
and $w_{s}$, which is the weight factor. The utility of such an approach comes from the fact that the parameters defined by the IGM framework will have a slightly different impact at the various stages of the EoR (for example, the role of cross terms on the total 21-cm power spectrum), while also allowing us to choose regions of the data with tighter measurements when considering mock data with more realistic noise estimates. We set $w_s = 1$ at all redshifts in this work, retaining the weighting for future application to data where some redshift bins are measured more tightly than others. 

In general, we initiate a number of walkers equal to $32\times N_{\mathrm{dim}}$, where $ N_{\mathrm{dim}}$ is the number of our evolution parameters. We ensure a burn-in stage of 20 per cent of our total run to pre-sample the grid, and run chains with up to $10^6$ steps. To quantify the convergence, we compute the Gelman–Rubin statistic $\hat{R}$ \citep{Gelman1992}, the ratio of between-chain to within-chain variance, which approaches 1 for converged chains. All parameters satisfy $\hat{R} -1 <10^{-2}$, indicating strong convergence. We also track the integrated autocorrelation time $\tau_{\mathrm{t}}$ of each parameter chain to ensure the total number of steps greatly exceeds $\tau_{\mathrm{t}}$ such that retained samples are effectively independent. We compute the global $\chi^{2}$ as the sum of squared residuals over all snapshots and $k$-bins of $\log_{10}|P_{21}/P_{\rm \delta_{\rho} \delta_{\rho}}|$, assuming a uniform log-space scatter $\sigma$, as well as with the $R^{2}$ metric, which measures the fraction of the variance in the ratio captured by the model. In Appendix~\ref{sec:forecast}, we use our method to generate a forecast study in order to test the constraining ability of our method.
\section{Results}
\label{sec:results}
We have developed a model of the 21-cm power spectrum, driven by five IGM parameters shown in Eq.~\eqref{eq:p21_model}. The expression is physically informed and redshift-independent, and is similar to phenomenological forms found within the literature \citep{Mirocha2022,Ghara2024}. To handle observations across reionization, we further expand the framework by providing functional forms for the evolution of the five IGM parameters in Sec.~\ref{sec:IGM_params}, making the model redshift-dependent. Following the IGM-centred approach, we select simple, physically informed parametrisations. We refer to these as evolution parameters to distinguish them from the IGM parameters. We present the results of an MCMC analysis (Sec.~\ref{sec:mcmc}), run with the physically constrained priors in Table~\ref{tab:priors} on 21-cm power spectra from {\sc C$^{2}$-Ray} (\texttt{C2R}), {\sc 21cmFAST} (\texttt{21F}), and {\sc POLAR} (\texttt{PLR}), the details of which are outlined in Sec.~\ref{sec:sims}.
\begin{figure*}
    \centering
    \includegraphics[width=1.0\linewidth]{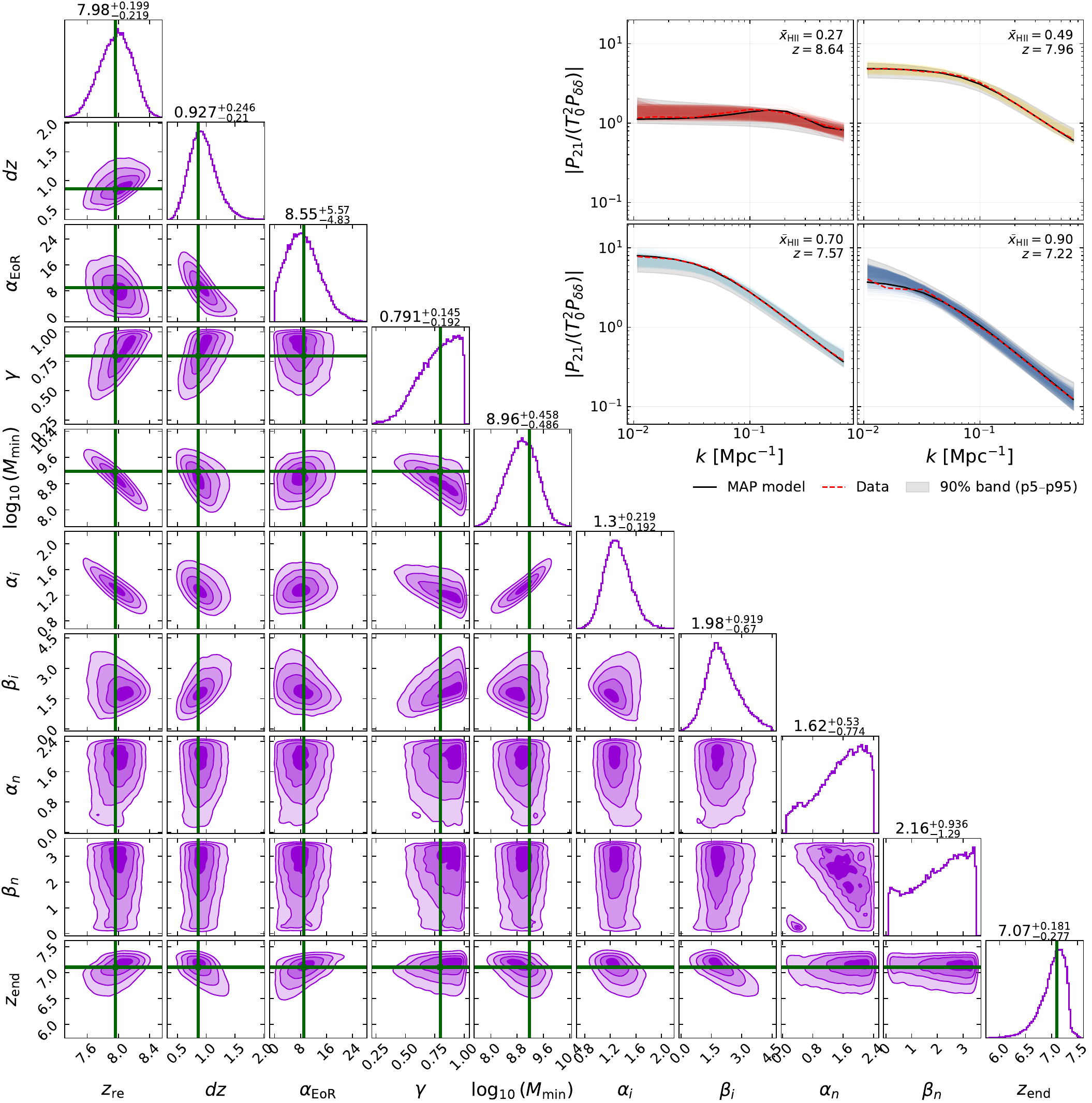}\caption{Posterior distributions of the evolution parameters (in increments of $0.5\sigma$) used within the IGM framework for the C$^{2}$-Ray simulation \texttt{C2R}, which are combined to produce the IGM parameters seen in Fig.~\ref{fig:C2hist}. The green lines represent the true values and each evolution parameter inferred directly from the global mass- and volume- weighted ionisation histories (see Sec.~\ref{sec:IGM_params}) whilst the minimum mass of halos hosting ionising sources $M_{\mathrm{min}} = 10^{9} M_{\odot}$ is known a priori from Sec.~\ref{sec:C2-Ray}. Upper right corner: Ratio of the 21-cm power spectrum to the matter density power spectrum, inferred from the posterior distributions. The true spectra are shown as a red dashed line, the best model is shown as a black line, a selection of good $\chi^{2}$ models are shown in the coloured regions, while the shaded regions show the sampled parameter space. We see that the data is well described by the IGM framework. Note that small differences between the two can be observed at the final ionisation slice in the furthest right panel of the second row, where $\xa = 0.90$ for $k< 0.03$ cMpc$^{-1}$. Rather than a physical property of the EoR, this is an unphysical feature which appears when applying a sharp barrier of $R_{\mathrm{max}} = 70$ cMpc \citep[see][]{Georgiev2022}.} 
    \label{fig:c2ray-posterior}
\end{figure*}
\subsection{\texorpdfstring{C$^{2}$-Ray model, \texttt{C2R}}{C2-Ray model, C2R}}
\label{sec:results_c2ray}
\begin{figure*}
    \centering
    \includegraphics[width=1.\linewidth]{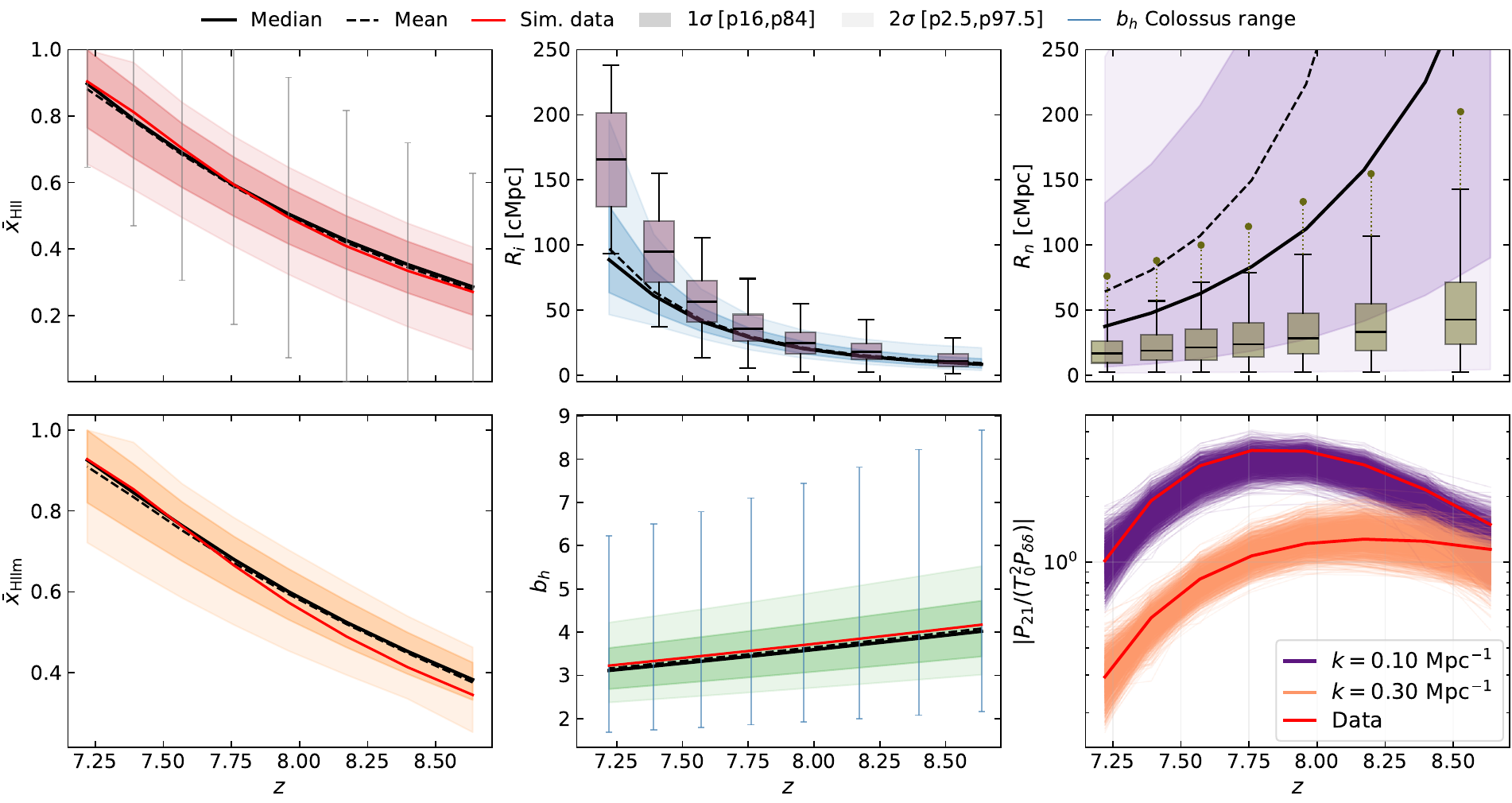}
    \caption{IGM framework parameters from Eq.~\eqref{eq:p21_model} derived from the posterior distributions shown in Fig.~\ref{fig:c2ray-posterior} for the {\sc C$^{2}$-Ray} model described in Sec.~\ref{sec:C2-Ray}. When available, the true value is shown as a thick red line, while the posterior mode, median, and mean are shown as thin dashed and dotted black lines, respectively. The shaded regions show the 1 and 2$\sigma$ high-density regions around the median, better capturing the probability distribution. Uncertainties on the volume- and mass-averaged ionisation fractions $\xa$ and $\xc$ are shown as light bars and are directly sampled from the simulation outputs. Additionally, the bars in the second and third panels of the first row show the 1$\sigma$ distributions of the ionised and neutral region sizes, $R_i$ and $R_n$, estimated with the RMFP-BSD method from Georgiev et al. (2026b, submitted). The true halo bias value from the simulation is shown against an uncertainty estimated by drawing samples from the posterior and assuming either a sampling or modelling uncertainty using different parametrisations of halo bias, such as \citet{Tinker2010, Sheth2001}. The final panel shows the true 21-cm signal ratio versus the density field at different redshifts, using a selection of sample models generated with the framework.} 
    \label{fig:C2hist}
\end{figure*}
We discuss the results of the framework on the {\sc C$^{2}$-Ray} model outlined in Sec.~\ref{sec:C2-Ray}. The MCMC analysis was first run with a burn-in stage of 20 per cent of the total run, after which it was run for $10^{6}$ steps to ensure convergence of the posterior distributions. We calculate the global statistical parameters such as the global $\chi^{2} \approx 2.59$ and the $R^{2} \approx 0.987$ goodness of fit. We present the posterior distribution corner plot in Fig.~\ref{fig:c2ray-posterior} and the evolution of the IGM parameters shown in Fig.~\ref{fig:C2hist}. In general, we find the evolution parameters well constrained and consistent with the model expectations, while the IGM parameters generally follow the values measured from the simulation. For the remainder of the section, we expand in detail on the inherent degeneracies between the parameters and the applicability of the framework across the different stages of the EoR.

We firstly focus on the top three rows of Fig.~\ref{fig:c2ray-posterior}, which showcase the global ionisation history parameters discussed in Sec.~\ref{sec:ion_history}. The framework constrains the ionisation history well, as shown in the first panel of Fig.~\ref{fig:C2hist}. We observe an inherent degeneracy between $\alpha_{\mathrm{EoR}}$ and $dz$ in the third row, second column of Fig.~\ref{fig:c2ray-posterior}. Taking the log of Eq.~\eqref{eq:reion_parametrisation} shows an anti-correlation between $\alpha_{\mathrm{EoR}}$ and $dz$ (see Fig.~\ref{fig:forecast_correlation} for the Pearson correlation coefficient between the two parameters). Intuitively, models with a shorter overall duration of reionization but a rapid increase could be degenerate with model histories with a longer duration of the EoR and shallower evolution of $\alpha_{\mathrm{EoR}}$. Part of this behaviour is imposed by the parametrisation rather than the data, since $z_{\rm re}$ and $dz$ enter the shape function $t(z)$ of Eq.~\eqref{eq:reion_parametrisation} only through the difference $z_{\rm end} = z_{\rm re} - dz$, and are separated only by the normalisation at the midpoint.

We shift focus to $M_{\mathrm{min}}$ and $\gamma$, shown in the fourth and fifth row of Fig.~\ref{fig:c2ray-posterior}. Although we sample for the minimum mass capable of hosting ionising sources $M_{\mathrm{min}}$, this parameter enters the IGM interpretation through its role in dictating the large-scale halo bias described in Sec.~\ref{sec:clustering}. It effectively captures the correlation between the halo field (and hence the density field) and the ionisation field. The logarithm of the minimum mass of reionization sources has been constrained to be $\log_{10}(M_{\mathrm{min}}) = 8.96^{+0.458}_{-0.486}$, the evolution of $b_{\mathrm{h}}$ is shown against redshift in the middle of the bottom row of Fig.~\ref{fig:C2hist}, matching that of the \texttt{C2R} simulation. Note, however, that $M_{\mathrm{min}}$ can exhibit a degeneracy with the midpoint of reionization $z_{\mathrm{re}}$ (fifth row, first column) as well as the ionisation bubble parameter $\alpha_{i}$ (sixth row, fifth column). The midpoint of reionization and the minimum mass are anti-correlated as both parameters influence the large-scale amplitude of the 21-cm power spectrum. While the halo bias directly impacts the amplitude, the midpoint for reionization coincides with the point of the largest variance of the field, when both ionised and neutral fluctuations are comparable \citep{Furlanetto2006a}. Therefore, models with a slightly higher minimum mass reach the EoR midpoint at a lower redshift, and vice versa. Additionally, $M_{\mathrm{min}}$ is positively correlated with $\alpha_{i}$, and therefore, the size of ionised regions ($R_{i}$) at the midpoint of reionization (see Eq.~\eqref{eq:Ri}). The growth of the characteristic scale of ionised regions will impact the amplitude of the 21-cm power spectrum through the transfer function in Eq.~\eqref{eq:p21_model} such that a larger $R_i$ will effectively decrease the amplitude of the transfer function as $T \propto 1/\sqrt{1+(kR_i)^{2}}$. However, degeneracy between the two parameters can be significantly reduced by informing the priors on the ionisation bubble parameters with measurements of ionised bubble sizes. Lastly, the correlation parameter $\gamma$ (relating $ \xc = \xa^{\gamma}$), although prior-limited, is relatively well constrained. The true value shown in the fourth row of Fig.~\ref{fig:c2ray-posterior} is obtained by directly fitting the two global fractions, yielding $\gamma \approx 0.791$. 

We discuss the ionisation morphology parameters in the final four rows of Fig.~\ref{fig:c2ray-posterior}. The $\alpha$ parameter describes the size of the region at a specified point of reionization, while $\beta$ governs the growth rate for the ionised regions and neutral patches, respectively. The ionisation morphology parameters $\alpha_{i}$ $\&$ $\beta_{i}$ are well constrained by the MCMC, and the resulting estimate of $R_i$ in Fig.~\ref{fig:C2hist} matches the evolution of the RMFP-BSD method from Georgiev et al. (2026b, submitted). For consistency, we retain the MFP-BSD box plots for $R_i$ above $\xa > 0.85$ and for $R_n$ below $\xa < 0.5$, even though these regimes contribute little to the power spectrum shape. As $\alpha_{i}$ and $\beta_{i}$ also affect the amplitude of the 21-cm power spectrum through their role in the transfer function, there naturally exists a slight correlation with $M_{\mathrm{min}}$ and $z_{\mathrm{re}}$, which is significantly reduced by the fact that the fit of the ionisation morphology parameters constrains both the amplitude and $k$-scale dependence of the power spectrum. Note, however, that each of these parameters is more difficult to constrain for values of $R_i$ such that the characteristic bubble scale becomes comparable to and smaller than $k\le 0.1$ cMpc$^{-1}$, a regime where the uncertainty on the 21-cm measurement is increased due to foreground residuals and sample variance. Conversely, the neutral patch parameters at the bottom right rows of Fig.~\ref{fig:c2ray-posterior} remain poorly defined, yet the overall evolution seen in Fig.~\ref{fig:C2hist} is generally recovered. The difficulty in constraining $R_n$ stems from the $\xa$ parametrisation, as the asymmetric power-law evolution in Eq.~\eqref{eq:reion_parametrisation} favours EoR models that rapidly re-ionise in their final stages. It is precisely in this regime that the neutral component of our IGM framework is most constraining, since the ionisation/neutral field is then dominated by neutral islands. Moreover, as the evolution of $\xa$ in Eq.~\eqref{eq:reion_parametrisation} ceases past the saturation limit, $R_n$ growth freezes in the final stages of the EoR. Nevertheless, the constraints of Fig.~\ref{fig:C2hist} are informative due to the constraining power of the priors placed on $\alpha_{n}$ $\&$ $\beta_{n}$.

Lastly, we inspect the true 21-cm power spectra in the top-right panel of Fig.~\ref{fig:c2ray-posterior}. The grey bands show the spread around the distribution mode. The thick black line represents the best-fit model, while the thin coloured lines represent a random selection of models with low $\chi^{2}$ statistics. The parametrisation describes the data very well at a given redshift as well as across reionization, with the best-fitting models lying within the assumed uncertainty. Because the IGM framework parameters are degenerate, the posterior distributions in Fig.~\ref{fig:c2ray-posterior} are more skewed, and the resulting model spread is not evenly distributed across each $k$-bin. Additionally, the parametrisation cannot match the low-$k$ features seen in the final stages of the data. The fluctuations in the data around $k \approx 0.03$ cMpc$^{-1}$ correspond to the hard barriers imposed on ionising photons at $R_{\mathrm{max}} = 70$ cMpc. This results in an unphysical drop in power \citep[see][]{Georgiev2022} and is more a property of the simulation model than of the EoR.

In summary, the IGM framework performs well on \texttt{C2R}, recovering the mass- and volume-averaged global ionisation history and the clustering of ionising sources. We infer the evolution of ionised regions from the 21-cm power spectra during the EoR and find it agrees well with the RMFP-BSD method. While we find similar results for the evolution of neutral islands, the neutral component of our framework is most constraining in the final stages of the EoR, where the parametrisation saturates, and the evolution of $R_n$ freezes.
\subsection{21cmFAST model}
\label{sec:results_21cmFAST}
\begin{figure}
    \centering
    \includegraphics[width=\columnwidth]{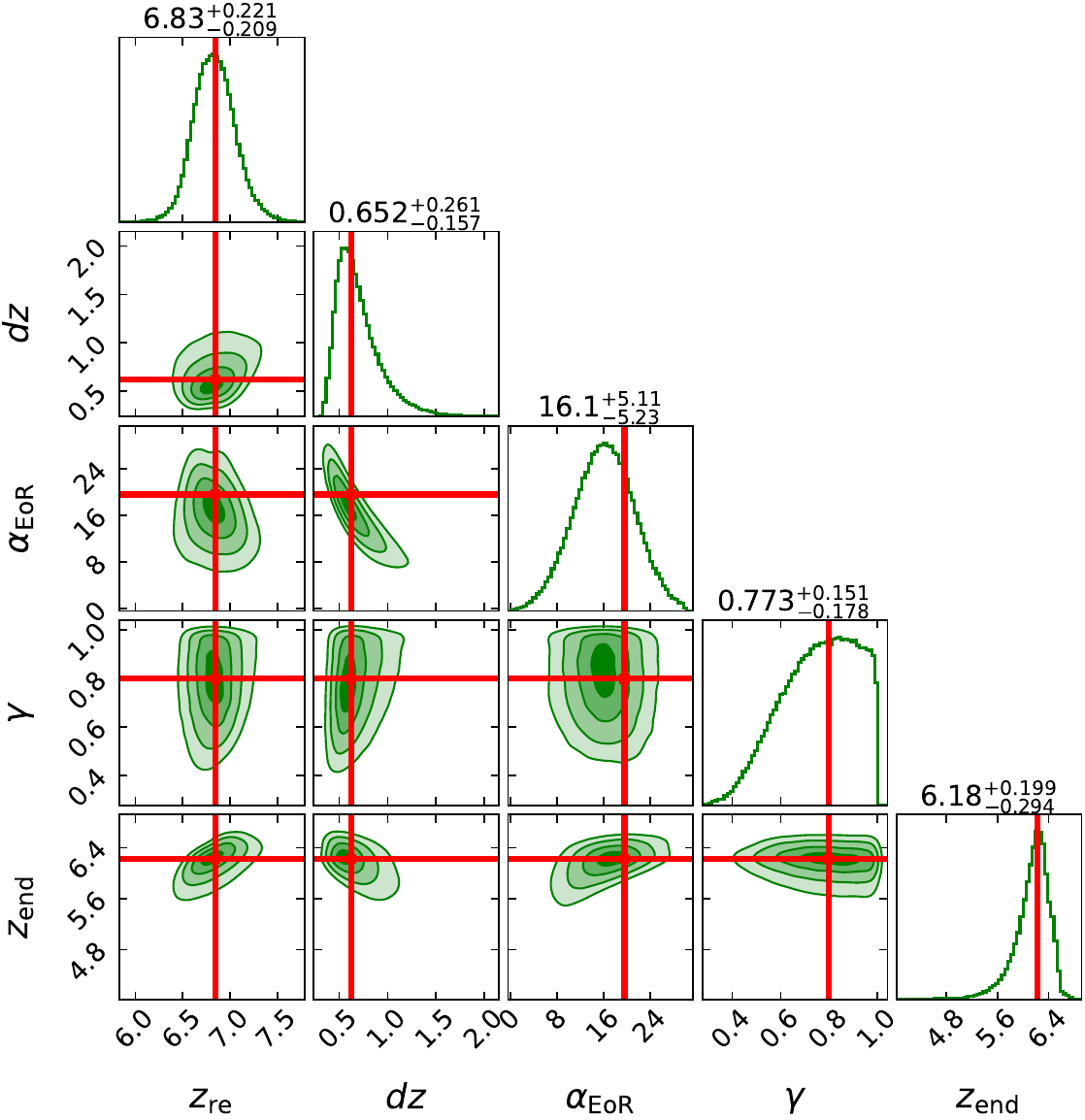}
    \includegraphics[width=\columnwidth]{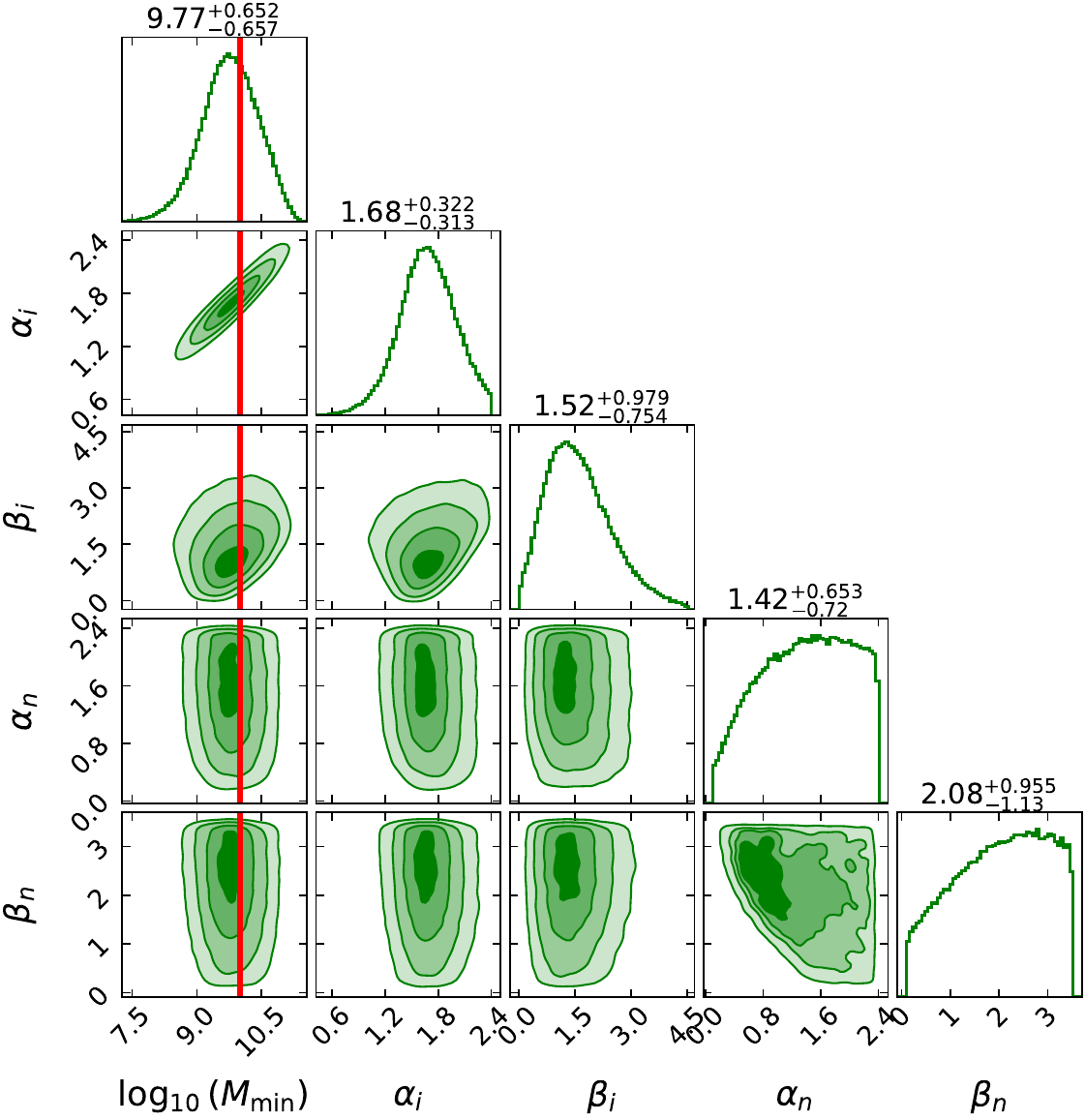}
    \caption{Posterior distributions of the evolution parameters used in the IGM framework for the {\sc 21cmFAST} simulation, combined to produce the IGM parameters shown in Fig.~\ref{fig:21cmFAST_IGM_params}.}
    \label{fig:21cmFAST_posteriours}
\end{figure}

\begin{figure*}
    \centering
    \includegraphics[width=1.\linewidth]{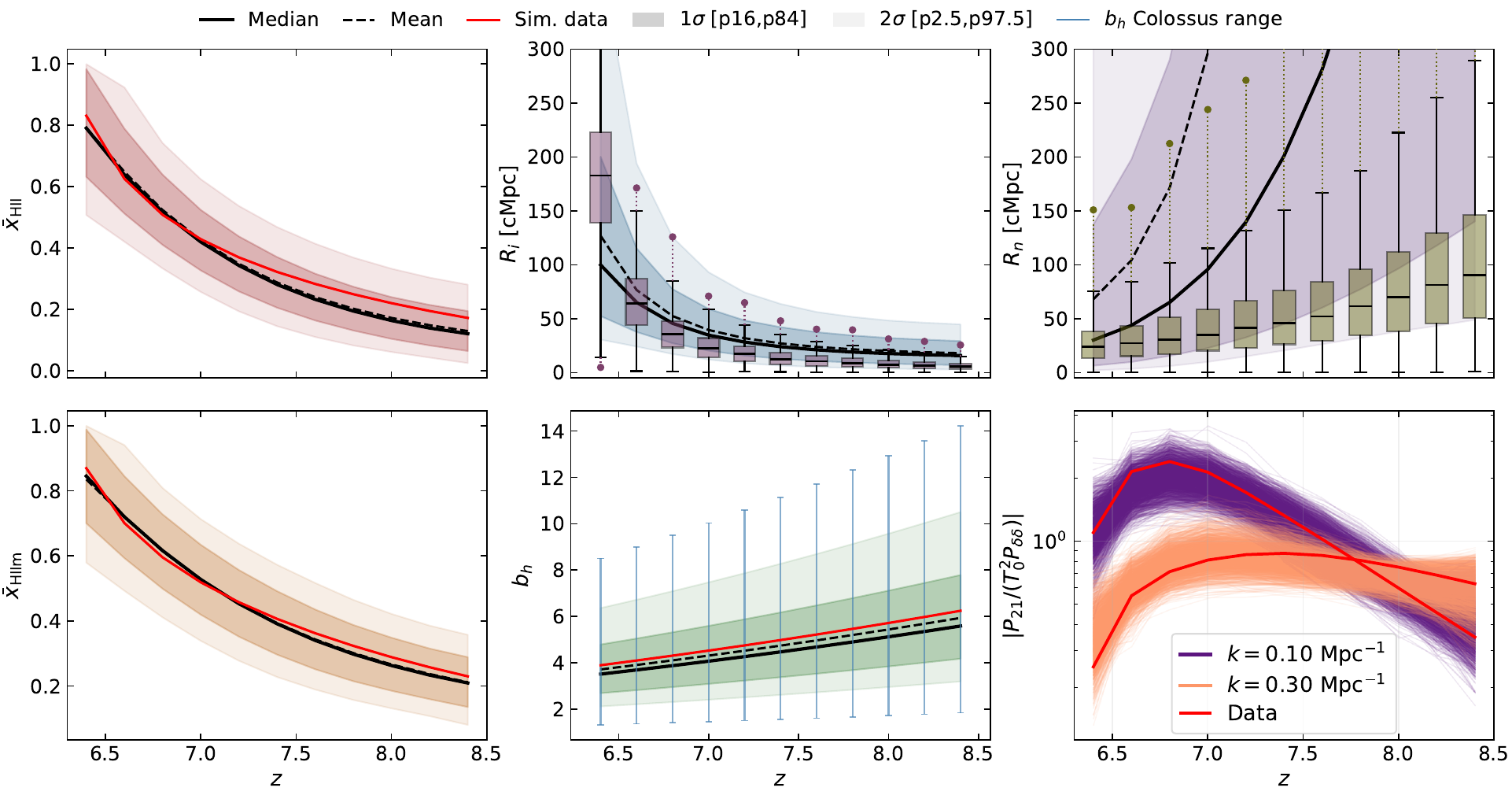}
    \caption{IGM framework parameters derived from the posterior distributions shown in Fig.~\ref{fig:21cmFAST_posteriours} for the \texttt{21F} model, arranged as in Fig.~\ref{fig:C2hist}.}
    \label{fig:21cmFAST_IGM_params}
\end{figure*}

We also test the IGM framework on the \texttt{21F} model outlined in Sec.~\ref{sec:21cmFAST}. We apply the prior ranges reported in Table~\ref{tab:priors}, running the analysis until the chain has converged and the $\hat{R}$ test is satisfied, with a global value of $\chi^{2} = 1.866$ and global $R^{2} = 0.974$. We visualise the posterior distributions of the evolution parameters in Fig.~\ref{fig:21cmFAST_posteriours} and their corresponding IGM parameters in Fig.~\ref{fig:21cmFAST_IGM_params}. 
\begin{figure}
    \centering
    \includegraphics[width=\columnwidth]{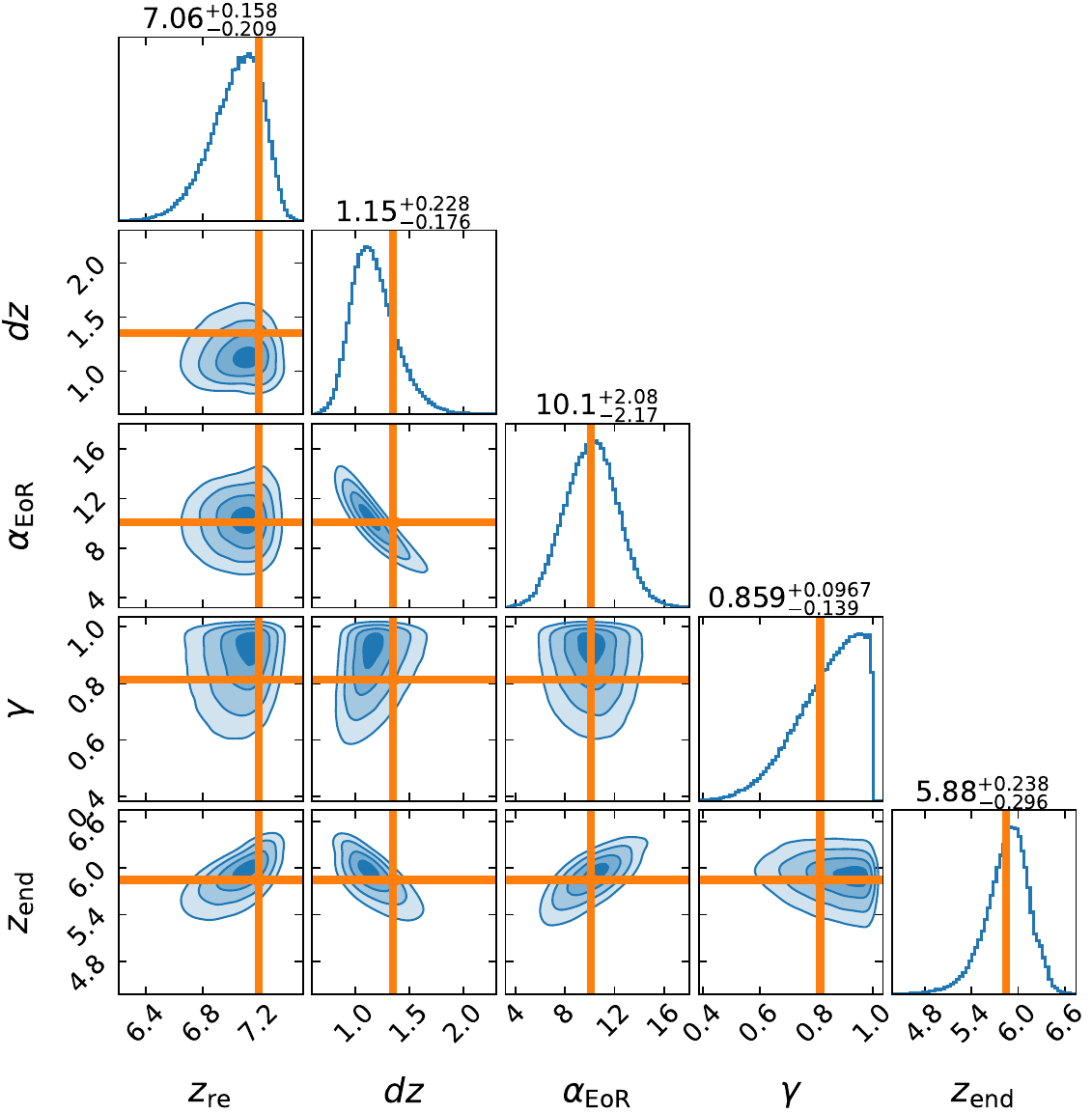}
    \includegraphics[width=\columnwidth]{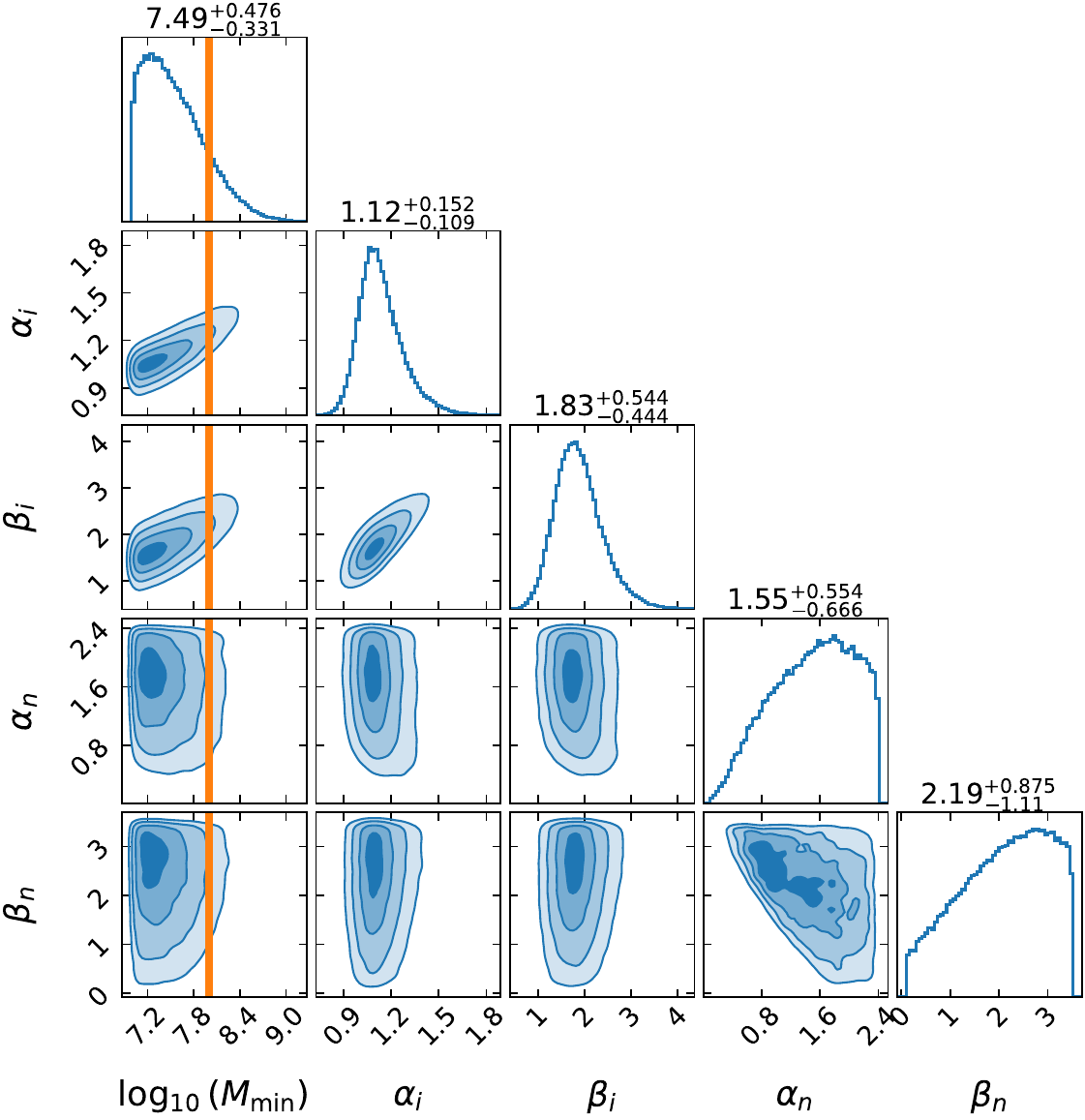}
    \caption{Posterior distributions of the evolution parameters used in the IGM framework for the {\sc POLAR} simulation, combined to produce the IGM parameters shown in Fig.~\ref{fig:polar_igm_params}.}
    \label{fig:polar-posterior}
\end{figure}
\begin{figure*}
    \centering
    \includegraphics[width=1.\linewidth]{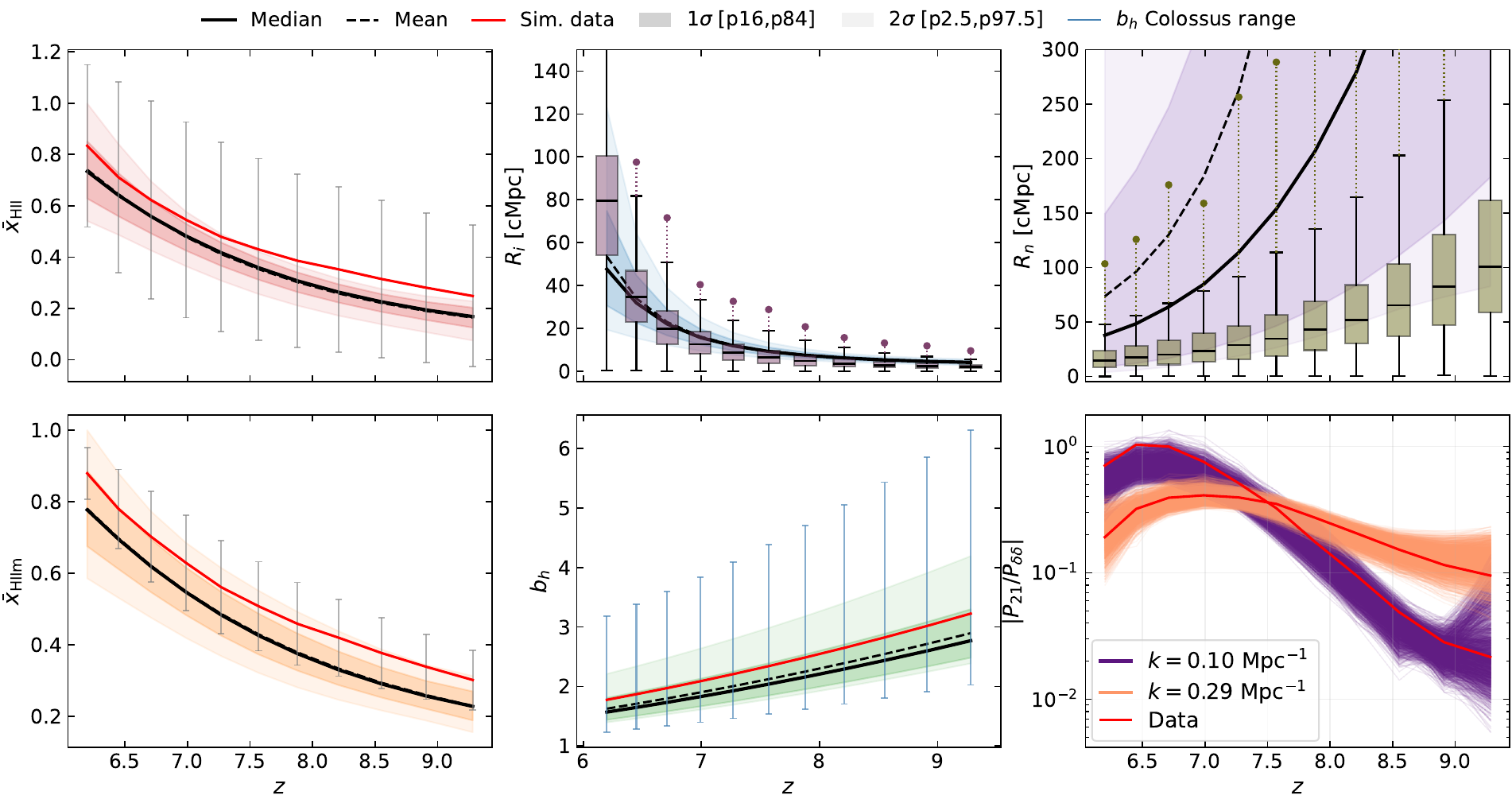}
    \caption{IGM framework parameters derived from the posterior distributions shown in Fig.~\ref{fig:polar-posterior} for the \texttt{PLR} model, arranged as in Fig.~\ref{fig:C2hist}.}
    \label{fig:polar_igm_params}
\end{figure*}
The global volume-weighted ionisation history parameters ($z_{\mathrm{re}}$, $dz$, $\alpha_{\mathrm{EoR}}$), shown in the first three rows of Fig.~\ref{fig:21cmFAST_posteriours} are well constrained. However, a degeneracy between $dz$ and $\alpha_{\mathrm{EoR}}$ is evident, with a slight tail of $dz$ toward higher values and a 1$\sigma$ offset in $\alpha_{\mathrm{EoR}}$. One way to understand this is to consider the choice of the power-law $\xa$ parametrisation in Eq.~\eqref{eq:reion_parametrisation} compared to the visibly exponentially evolving global ionisation history shown in the first panel of Fig.~\ref{fig:21cmFAST_IGM_params}. However, in this simulation, with fewer redshift data points than in the \texttt{C2R} model, the degeneracy between the parameters can become more pronounced. The clustering parameter $b_h$ is well constrained with respect to the posterior distribution (first column in the bottom panel of Fig.~\ref{fig:21cmFAST_posteriours}) and its redshift evolution, shown in the middle panel of the bottom row of Fig.~\ref{fig:21cmFAST_IGM_params}. Partly, this is a consequence of the excursion-set-like methodology of the {\sc 21cmFAST} code, which more tightly couples the ionisation morphology to the matter density field, likely resulting in a stronger clustering feature. Nevertheless, much of the constraining power comes from the large minimum mass of halos hosting ionising sources (in this case $M_{\mathrm{min}} = 10^{10} M_{\odot}$), which will result in a larger clustering effect on the 21-cm power spectrum amplitude. Lastly, the ionisation region parameter $R_{i}$ is well constrained by the posterior, and in Fig.~\ref{fig:21cmFAST_IGM_params} the prediction follows the RMFP-BSD data points quite well up to the very final phases of the EoR, when the power spectrum is instead determined by the neutral island morphology.

In summary, we conclude that the IGM framework is applicable to semi-numerical models such as {\sc 21cmFAST}, despite differences in the source parametrisation, clumping model, RT method, and density field. Because we have fewer power spectra data points, the natural degeneracies in the parametrisation of $\xa$ can become more pronounced when constraining the evolution parameters $z_{\mathrm{re}}$, $dz$, and $\alpha_{\mathrm{EoR}}$. Nevertheless, the joint posterior of $\xa$ remains well constrained, and in such cases the median of the distribution will be naturally more informative, as it is less sensitive to such degeneracies and outliers in the data.
\subsection{{\sc POLAR} model}
\label{sec:polar_results}
The {\sc POLAR} simulation used in this study, referred to as \texttt{PLR} for simplicity, is an important stress test for the IGM framework, helping us investigate theoretical and numerical assumptions that the \texttt{C2R} and \texttt{21F} models cannot address. Firstly, the simulation box is smaller than the aforementioned models, covering 24 per cent of their volume, and may fail to capture rare environments that might impact the course of reionization \citep{Iliev2014}. To limit the effects of sample variance, we sample the power spectrum with fewer $k$-bins to average out some of the noise. We remove $k$-scales below $k = \frac{2 \pi}{0.5L} \approx 0.03$ cMpc$^{-1}$, corresponding to modes larger than half of the simulation's box size. We also apply a high-$k$ cut-off to account for aliasing effects. The low-$k$ cut-off will make it difficult to constrain neutral/ionised regions above 50 cMpc. Moreover, this data set contains only around ten redshift bins during the EoR, fewer than in the previous analyses. We therefore expect larger uncertainties on the recovered parameters.

The analysis of the data set derives a global $\chi^{2} \approx1.986$ and $R^{2} \approx 0.986$, and each iteration is run to convergence. As in previous sections, the posterior distribution of the evolution parameters is shown in Fig.~\ref{fig:polar-posterior}, the IGM parameters as they evolve across redshifts are seen in Fig.~\ref{fig:polar_igm_params}. In general, we report a well-constrained parameter space corresponding to the true values of the \texttt{PLR} model and the evolution of the IGM parameters consistent with the median of each reconstructed posterior distribution. The framework performs well despite the small box size, considering the implementation method of the properties of the ionisation sources and the method of radiative transfer.

We instead focus on the difficulty the framework faces when analysing the {\sc POLAR} data. We address the deviations in the evolution parameters, the constraints on the minimum mass of halos hosting ionising sources $M_{\mathrm{min}}$ as well as the midpoint and duration of reionization $z_{\mathrm{re}}$ $\&$ $dz$, which are within 1$\sigma$ of the true value (see Fig.~\ref{fig:polar-posterior}). For these parameters, the important key factor is the assumption of how tightly the ionisation regions trace the first ionising sources. The previously discussed \texttt{C2R} and \texttt{21F} models cannot resolve structure associated with the dark matter halos and, in this set-up, do not consider feedback or stochastic processes of their ionising sources. In the \texttt{C2R} and \texttt{21F} models, the ionising photons are assigned to the pre-defined halos within the simulation through a parametrised model \citep[see][]{giri2021measuring}. The \texttt{PLR} model has a more physically informed, and thus more stochastic, assignment model of ionising sources, which accounts for changes in the emissivity of ionising sources due to effects such as mergers of galaxies, supernovae, AGN feedback, the reincorporation of ejected gas, and bursts of star formation \citep{Henriques2020}. Moreover, compared to \texttt{C2R} and \texttt{21F}, where $\dot{N}_{\mathrm{ion}} \propto M_{\mathrm{h}}$, the \texttt{PLR} galaxy-halo connection follows $M_{\mathrm{gal}} \propto M_{\mathrm{h}}^{1.5}$ (with a constant $f_{\mathrm{esc}} = 0.1$), hence, $\dot{N}_{\mathrm{ion}}$ is non-linear in $M_{\mathrm{h}}$. Because of this, it is not trivial to deduce whether the source model will increase or decrease the clustering of ionising sources and how it would impact the halo bias parameter $b_{\mathrm{h}}$.

Naïvely, we can build a generalised conjecture that ionising sources associated with low-mass atomically cooling halos (LMACH, $10^{8}M_{\odot} <M<10^{9}M_{\odot}$) will drive the formation of the ionisation morphology and be aided by bursts of star formation since they are substantially more numerous than high-mass atomically cooling halos (HMACH, $M>10^{9}M_{\odot}$) at the beginning of the EoR. Yet such sources are more prone to the various feedback processes, such as SN feedback, compared to HMACH \citep{Pawlik2015,Dixon2016}. Therefore, during the EoR, we would expect the larger atomically cooling halos to play a more active role in governing the clustering bias. Even though the analytic model used to infer $b_{\mathrm{h}}$ in the IGM framework in Eq.~\eqref{eq:halo_bias} will not be able to properly model such a shift in the halo bias's evolution, we expect most of the constraining power on the parameter to come from the 21-cm power spectrum at the beginning of reionization, as the clustering is of less importance after its midpoint. We can confirm this as the IGM framework's inferred minimum mass of the ionising sources of log$_{10} (M_{\mathrm{min}}) = 7.49^{+0.476}_{-0.331}$ in the fifth row of Fig.~\ref{fig:polar-posterior} is within 1$\sigma$ of the true value. The framework's preference for a slightly lower minimum mass of the halos hosting the ionising sources is understandable since mini-halos with masses of approximately $10^{7} M_{\odot}$ tend to trace the same structure as $10^{8} M_{\odot}$ halos \citep[see][]{Iliev2005}. Another way to explain this is to refer to the halo bias formula in Eq.~\eqref{eq:halo_bias} and note that differences between the inferred halo bias amplitude of the aforementioned minimum masses are around 10 per cent (as opposed to 50 per cent when comparing the minimum masses of the \texttt{C2R} and \texttt{21F} models), within the uncertainty limit of the data \citep[also see][]{Reed2009}. 

Additionally, the framework's preference towards a slightly lower value of $z_{\mathrm{re}}$ and $dz$, as seen in the first row of Fig.~\ref{fig:polar-posterior}, is also affected by the relationship between the ionisation and the density field. In short, an EoR model with source suppression such as \texttt{PLR} affects the coupling strength of the cross terms from Eq.~\eqref{eq:Decomposition}, shifting the point of equilibration and the sign flip of the cross terms towards a slightly lower neutral fraction and, hence, a slightly lower redshift. We discuss this in detail in Appendix~\ref{sec::polar_decomposition}.

To summarise, the framework can be applied to the \texttt{PLR} model to infer the IGM parameters influencing the EoR, retrieving the volume- and mass-weighted ionisation histories as well as the properties of the neutral and ionised regions across the EoR (Fig.~\ref{fig:polar_igm_params}). Most interesting for this discussion are the constraints on $M_{\mathrm{min}}$, $z_{\mathrm{re}}$, and $dz$ (Fig.~\ref{fig:polar-posterior}): while constrained within 1$\sigma$ of the truth, all three are slightly biased towards lower values. This bias is directly linked to the 21-cm ratio in Fig.~\ref{fig:polar-p21-spectra}, where the signal power at low-$k$ is lower than expected from the prediction in Eq.~\eqref{eq:transfer_function}. This is due to the {\sc L-GALAXIES} source model \citep{Henriques2020}, whose feedback hinders the growth of ionised regions around LMACH early on, resulting in a patchier EoR morphology rich in partially ionised cells, which in turn more strongly couples the anti-correlation between the ionisation and density fields. The result is a shift in the epoch of equilibration \citep{Lidz2007,Georgiev2022}, as illustrated by the peak feature in Fig.~\ref{fig:polar_dec_terms}, reducing the low-$k$ power of the 21-cm signal and prolonging it into the latter part of the EoR, which the fit then compensates for by lowering the halo bias, as seen in the middle panel of the bottom row of Fig.~\ref{fig:polar_igm_params}. 
\section{Conclusions}
\label{sec:conclusions}
In this work, we have developed an IGM-centric model of the 21-cm power spectrum from the epoch of reionization, which is conceptually distinct from source-centric models. Since the 21-cm power spectrum is a direct probe of the intergalactic medium, our model has been built with the aim of describing the key aspects that dictate the reionization structure and morphology. This allows us to retrieve key reionization observables, which we refer to as the IGM parameters. These are the volume- and mass-averaged ionisation fractions ($\xa$, $\xc$) as well as the characteristic scale of ionised regions ($R_{i}$), linking the latter to the growth of ionised percolation clusters in the EoR. Since the IGM is a two-phase medium consisting of neutral and ionised hydrogen, our model also accounts for the role neutral islands ($R_{n}$) play in the ionisation power spectrum during the final stages of the EoR. Because the first ionising sources of reionization will originate in the densest regions of the Universe, we model the dependence on the matter density power spectrum. Moreover, we account for the clustering of the first ionising sources by including the source bias associated with the halo bias ($b_{\mathrm{h}}$) of DM halos. Lastly, we model the non-trivial yet non-negligible role of cross-terms highlighted in Eq.~\eqref{eq:Decomposition}, which capture the large-scale coupling between the density and ionisation/neutral field (see Appendix~\ref{sec:cross-terms}).

The IGM-centred framework can be applied to single redshift data as well as across a wide redshift range during the EoR. We achieve this by fitting the IGM parameters to various models across the literature, choosing functional forms which are also physically informed and complementary to other reionization observables (see Table~\ref{tab:priors}). We apply our framework to 21-cm power spectra generated by the {\sc C$^{2}$-Ray} (\texttt{C2R}), {\sc 21cmFAST} (\texttt{21F}), and {\sc POLAR} (\texttt{PLR}) simulations. These simulations cover a diverse range of EoR modelling techniques, from the method of radiative transfer to the properties of the ionising sources, as well as variations in cosmological parameters, box size, and gridding (see Sec.~\ref{sec:sims} for a detailed description). This allows us to investigate the performance of the IGM framework across a wide range of theoretical and computational assumptions.
We report the following findings.
\begin{enumerate}
    \item[1.] In the \texttt{C2R} simulation, we find that the IGM parameters are well constrained and match the true values (Fig.~\ref{fig:c2ray-posterior} and Fig.~\ref{fig:C2hist}). We recover the volume- and mass-weighted global ionisation fractions with high precision and can place tight constraints on source clustering. While the morphology of ionised/neutral patches cannot be extracted directly from the simulation, we instead sample it with a radial mean free path bubble size distribution (RMFP-BSD) method from Georgiev et al. (2026b, submitted). We find its estimates of the ionised and neutral region sizes agree with our constraints within our expectations.
    \item[2.] The \texttt{21F} simulation shares the same large box size and cell resolution as the \texttt{C2R} model but has been generated with the semi-numerical EoR code {\sc 21cmFAST}. It applies a different source parametrisation \citep{Davies2025}. The IGM framework also recovers the evolution parameters (Fig.~\ref{fig:21cmFAST_posteriours}) and the IGM parameters (Fig.~\ref{fig:21cmFAST_IGM_params}). The global ionisation fraction in the first panel of the figure, while still well constrained within $1\sigma$, appears to slightly underestimate the global evolution for values of $\xa < 0.2$. Comparing the ionisation histories of the \texttt{C2R} and \texttt{21F} models, the latter re-ionises exponentially towards the end of the EoR. The asymmetric power law model from Eq.~\eqref{eq:reion_parametrisation} cannot fully capture this ionisation history, prioritising the later stages, resulting in a more pronounced anti-correlation between $dz$ and $\alpha_{\mathrm{EoR}}$ seen in the top panel (third row, second column) of Fig.~\ref{fig:21cmFAST_posteriours} (though also see Fig.~\ref{fig:forecast_correlation}). We can therefore conclude that, even with a mismatched model of $\xa$, we can still retrieve the true history within  1$\sigma$, but we must be mindful of the evolution parameters and their inherent degeneracies. Naturally, we can marginalise the evolution parameters, or improve the parametrisation.
    \item[3.] Lastly, we test the framework against a {\sc POLAR} simulation from \citet{Ma2026}, which has a smaller simulation volume and uses 1D RT \citep{Ghara2015a} to model the EoR. Our constraints on the evolution parameters in Fig.~\ref{fig:polar-posterior} once again show broad agreement with the true values, though note the biasing of $z_{\mathrm{re}}$, $dz$, and $M_{\mathrm{min}}$ towards lower values (yet within uncertainty). This may be linked to the fact that the \texttt{PLR} model's galaxy formation prescription introduces a more complex relation between ionisation morphology and ionising sources. The {\sc L-GALAXIES} model includes various galactic feedback mechanisms \citep{Henriques2020}, which can aid or hinder the growth of ionising bubbles depending on the halo mass to which the ionising sources are assigned. This produces a patchier early EoR, rich in partially ionised cells, which strengthens the large-scale anti-correlation between the density and neutral fields. The resulting deficit in low-$k$ power, visible in the 21-cm ratio in Fig.~\ref{fig:polar-p21-spectra}, limits the constraining power on $b_{\mathrm{h}}$ and $z_{\mathrm{re}}$. This shows that relating the source bias to the halo bias, while an essential first step in constructing the IGM framework, can be improved further.
\end{enumerate}
In conclusion, we find our IGM parametrisation to perform remarkably well across a diverse range of reionization models in the literature, with varying numerical and theoretical foundations. Its ability to constrain properties of the EoR, at a given redshift or across a redshift range, provides a powerful new method to examine the 21-cm power spectrum measurements and upper limits with radio interferometers. The IGM framework naturally lends itself to joint analysis with observables from CMB data as well as high-redshift galaxies and quasars. Moreover, the method is complementary to the source-based inference approach currently used in upper-limit papers. These complementary methods can further constrain shared parameters, such as the ionisation history, or link source properties to the evolution parameters of the IGM framework. 
\section*{Acknowledgements}
IG, SZ, BC, AKS acknowledge support from the Israel Science Foundation (grant no. 1388/24) and the GIF “German-Israeli Foundation for Scientific Research and Development''. L.V.E.K. acknowledges support from the European Research Council (ERC) under the European Union’s Horizon 2020 research and innovation programme, grant agreement No. 884760 (CoDEX). AA acknowledges financial support from the Simons Foundation for their current position as a ``Learning the Universe'' and BCCP Fellow. SKG acknowledges support from Olle Engkvist Stiftelse (grant no.
232-0238). We have utilised the following python packages for manipulating the simulation outputs and plotting results: \texttt{numpy} \citep{harris2020array}, \texttt{scipy} \citep{2020SciPy-NMeth}, matplotlib \citep{Hunter2007} and {\sc tools21cm}\footnote{\url{https://github.com/sambit-giri/tools21cm}} package \citep{Giri2020}, \texttt{emcee} \citep{Foreman-Mackey2013}, \texttt{COLOSSUS} \citep{Diemer2018}.

\section*{Data Availability}
The data underlying this article will be shared upon reasonable request
to the corresponding author.



\bibliographystyle{mnras}
\bibliography{example} 




\appendix
\section{Cross contributions of the 21-cm power spectrum}
\label{sec:cross-terms}
The decomposition of the 21-cm power spectrum in Eq.~\eqref{eq:Decomposition} results in the derivation of cross spectra that describe the complex coupling between the ionisation and density fluctuations. Although non-trivial to model analytically, these terms have been shown to play an important role in dictating the shape and amplitude of the 21-cm power spectrum \citep[see][]{Lidz2007,Georgiev2022}. In this section, we model these terms and compare our outputs with simulations from Sec.~\ref {sec:sims}.

\begin{figure}
	\includegraphics[width=\columnwidth]{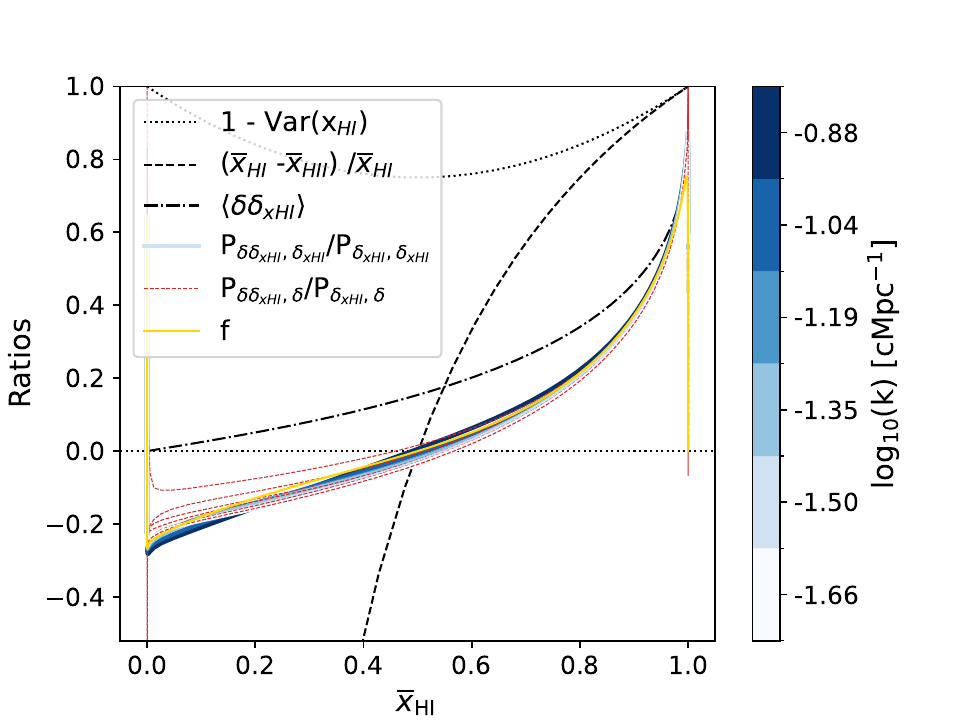}
    \includegraphics[width=\columnwidth]{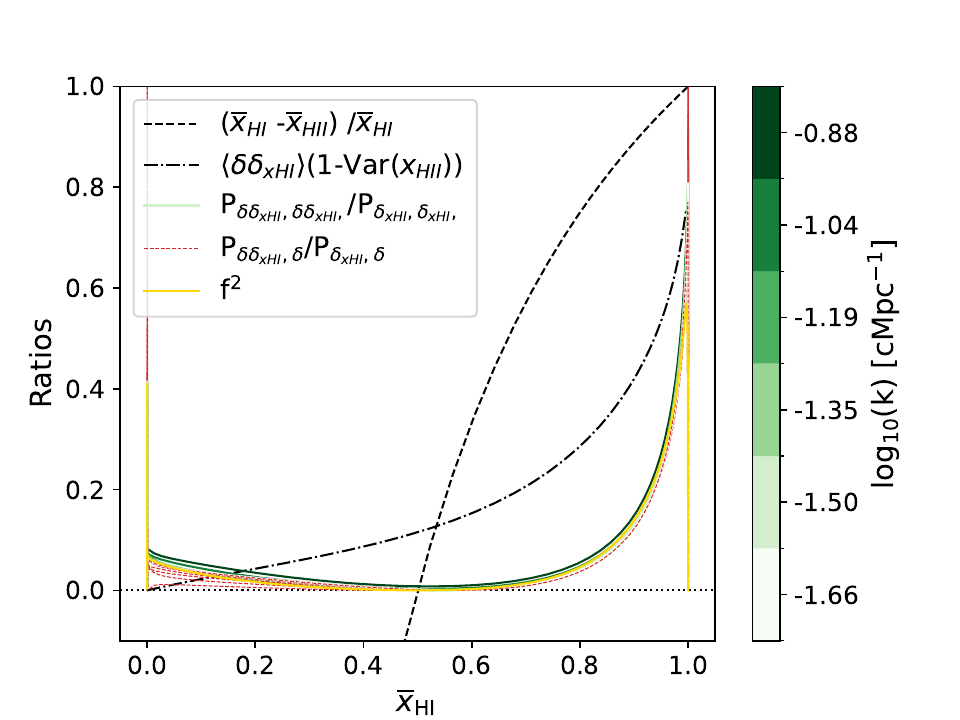}
    \caption{Ratios of power spectra from Eqs.~\eqref{fig:f} from the \texttt{C2R} simulation, described in Sec.~\ref{sec:C2-Ray}. These panels capture the large-scale behaviour of the cross terms in Eq.~\eqref{eq:Decomposition}. The top panel shows the ratio of spectra in which the density ionisation coupling parameter $f$ occurs once according to theory, while the bottom panel shows ratios of terms where it is squared. The thin black lines visualise the evolutions of the components of Eq.~\eqref{eq:f}, where the thin dotted line is the mode coupling from SPT between the density and ionisation fields, the dash-dotted line captures their large-scale correlation, while the dashed line models the sign flip, which occurs between the fluctuations of the two fields before and after the midpoint of reionization. The thick yellow line is the full model, which is seen to properly match the data throughout the EoR.}
    \label{fig:cov_approx}
\end{figure}

To better understand the relation between the cross terms and the auto- and cross-correlation of the ionisation field, we construct the following ratios, which we present in Fig.~\ref{fig:cov_approx}. For the lowest-$k$, we note the following relationship
\begin{equation}
\label{fig:f}
     f = \frac{P_{\delta_{\xHI} \delta, \delta}}{P_{\delta_{\xHI},\delta}} \approx \frac{P_{\delta_{\xHI} \delta, \delta_{\xHI}}}{P_{\delta_{\xHI},\delta_{\xHI}}},\quad f^{2} = \frac{P_{\delta_{\xHI} \delta, \delta_{\xHI} \delta}}{P_{\delta_{\xHI},\delta_{\xHI}}}
\end{equation}
The ratios for the spectra in the top panel of Fig.~\ref{fig:cov_approx} are remarkably similar throughout the reionization, starting at zero when the ionisation fluctuations are non-existent and quickly rising with the formation of the first sources. The function then decreases as the EoR progresses. On the one hand, this behaviour is governed by the mean fluctuation of $\langle \delta_{\xHI} \delta \rangle = (\xmb/\xb -1) = \mathrm{Cov}(\xHI,\delta)/\xb$ and arises from the large-scale covariance between the two fields \citep[see sec. 3.2 of][]{Lidz2007}. This expression is shown by the dash-dotted line for both panels of Fig.~\ref{fig:cov_approx} and indicates that the cross terms carry additional information about the large-scale coupling between the two fields. Furthermore, the function changes sign at the midpoint of reionization following the global change of the neutral perturbation \citep[see sec.~4.4 of][]{Georgiev2022}. Moreover, at the midpoint of the EoR, the universe transitions from a mostly neutral to a mostly ionised state, resulting in a sign flip for the higher-order fluctuations seen in fig. 2 of \citet{Georgiev2022}. To account for the sign change, we multiply the expression by a normalisation factor of $(\xb -\xa) /\xb$.  On the other hand, in modelling this function we must also take into account how the gravitational mode coupling of the small-scales of the cross-terms can propagate into the evolution of their spectra. Following Sec.~\ref{eq:SPT}, we can derive an additional term of $1-p\times \mathrm{Var}(\xHII)$, where $p$ is the resulting factor of the various terms in the mode coupling coefficient \citep[see sec. 3 of][]{Lidz2007}. This value ranges between $27/21$ for $P_{\delta_{\xHI} \delta, \delta}$  and $17/21$ for the $P_{\delta_{\xHI} \delta, \delta_{\xHI}}$ term, hence, we choose the mean value, which approximates to $1-\mathrm{Var}(\xHII)$. Lastly, the autocorrelation term $P_{\delta_{\xHI} \delta, \delta_{\xHI}, \delta}$ shown in the bottom panel of Fig.~\ref{fig:cov_approx}, which is a subdominant term throughout the EoR, can be approximated as the square of the function $f$. In summary, this allows us to converge upon the following form to account for the corrections to the large amplitudes of the cross-terms

\begin{equation}
\label{eq:f}
    f \approx \mathrm{Cov}(\xHI,\delta)  \Big(\frac{\xa - \xb}{\mathrm{Var}(\xHII)} \Big) (1-\mathrm{Var}(\xHII))
\end{equation}
In doing so, we can model each cross term as a function of Eq.~\eqref{eq:f} multiplied by either the cross or auto power spectrum of the ionisation field. Therefore, we can analytically remodel 
Eq.~\eqref{eq:Decomposition} using the true spectra and the model of $f$ as seen in Fig.~\ref{fig:cov_approx}. The results confirm that on low-$k$ scales the model can introduce a theory error of around 5 per cent, which can rise to about 10-20 per cent at the high-$k$ scales. 
\begin{figure}
\includegraphics[width=\columnwidth]{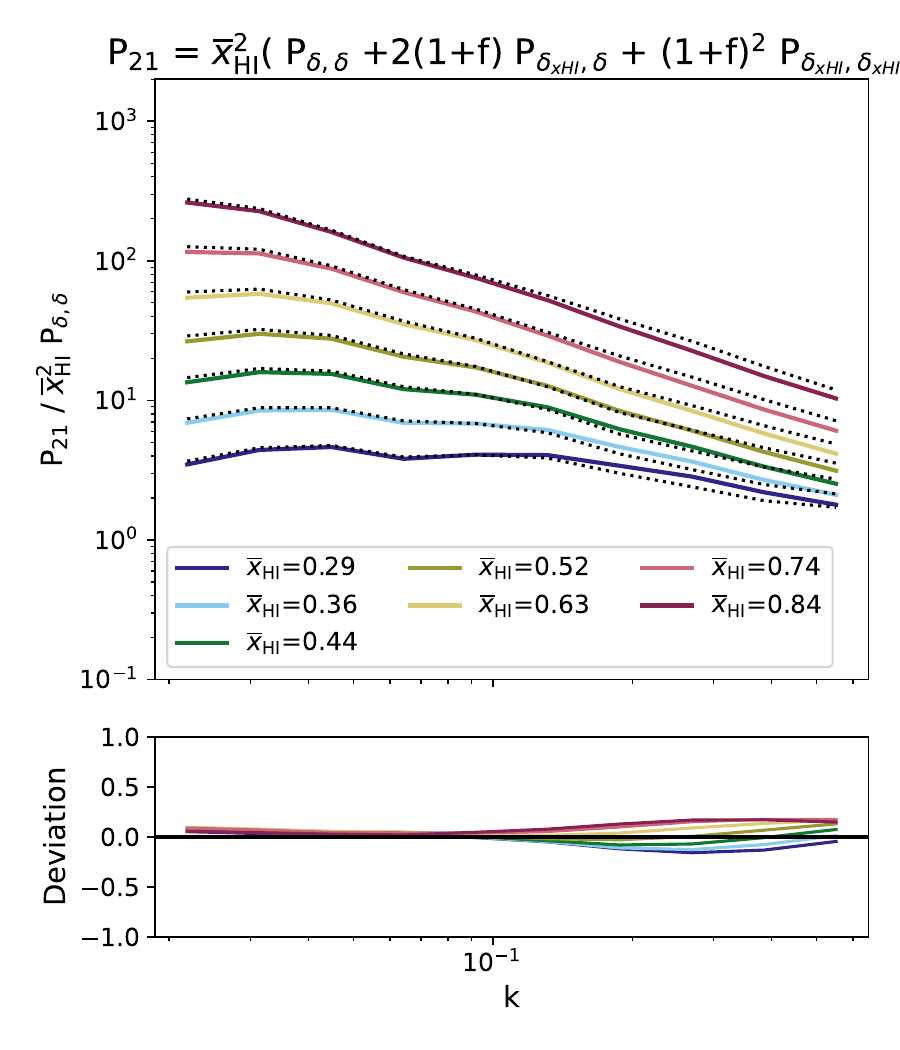}
    \caption{The 21-cm power spectrum throughout the EoR from the \texttt{C2R} simulation from Sec.~\ref{sec:C2-Ray} is shown in solid lines in the upper panel and compared to a simplified version of the 21-cm power spectrum where the cross terms present in Eq.~\eqref{eq:Decomposition} are modelled through the $f$ parameter derived in Eq.~\eqref{eq:f} and compared to the cross-term data in Eq.~\eqref{fig:f}. The bottom panel shows the theoretical error from the true values, highlighting small errors of less than 20 per cent during the EoR.}
    \label{fig:cov}
\end{figure}
\section{Forecast of the MCMC sample}
\label{sec:forecast}
In this section we conduct a forecast study of the IGM framework outlined in Sec.~\ref{sec:theory} using the MCMC sampler described in Sec.~\ref{sec:mcmc}. We focus on the analysis in Sec.~\ref{sec:results_c2ray} using the IGM framework on the \texttt{C2R} simulation outlined in Sec.~\ref{sec:C2-Ray}. To separate the natural degeneracies inherent to the sampler from the simulation data itself, we use the true values and best fits for \texttt{C2R} from Fig.~\ref{fig:c2ray-posterior} to generate our forecast. Some parameters like the minimum mass of halos hosting ionising sources and the midpoint of reionization are inherently discernible from inspection of the data. We directly infer parameters such as the duration, evolution, and density-ionisation parameters ($dz$, $\alpha_{\mathrm{EoR}}$, $\gamma$) from the true global ionisation history. Meanwhile, we choose the best-fit values for parameters related to the ionisation and neutral morphology using the updated MFP-BSD method from Georgiev et al. 2026b.
\begin{figure*}
    \centering
    \includegraphics[width=1.0\linewidth]{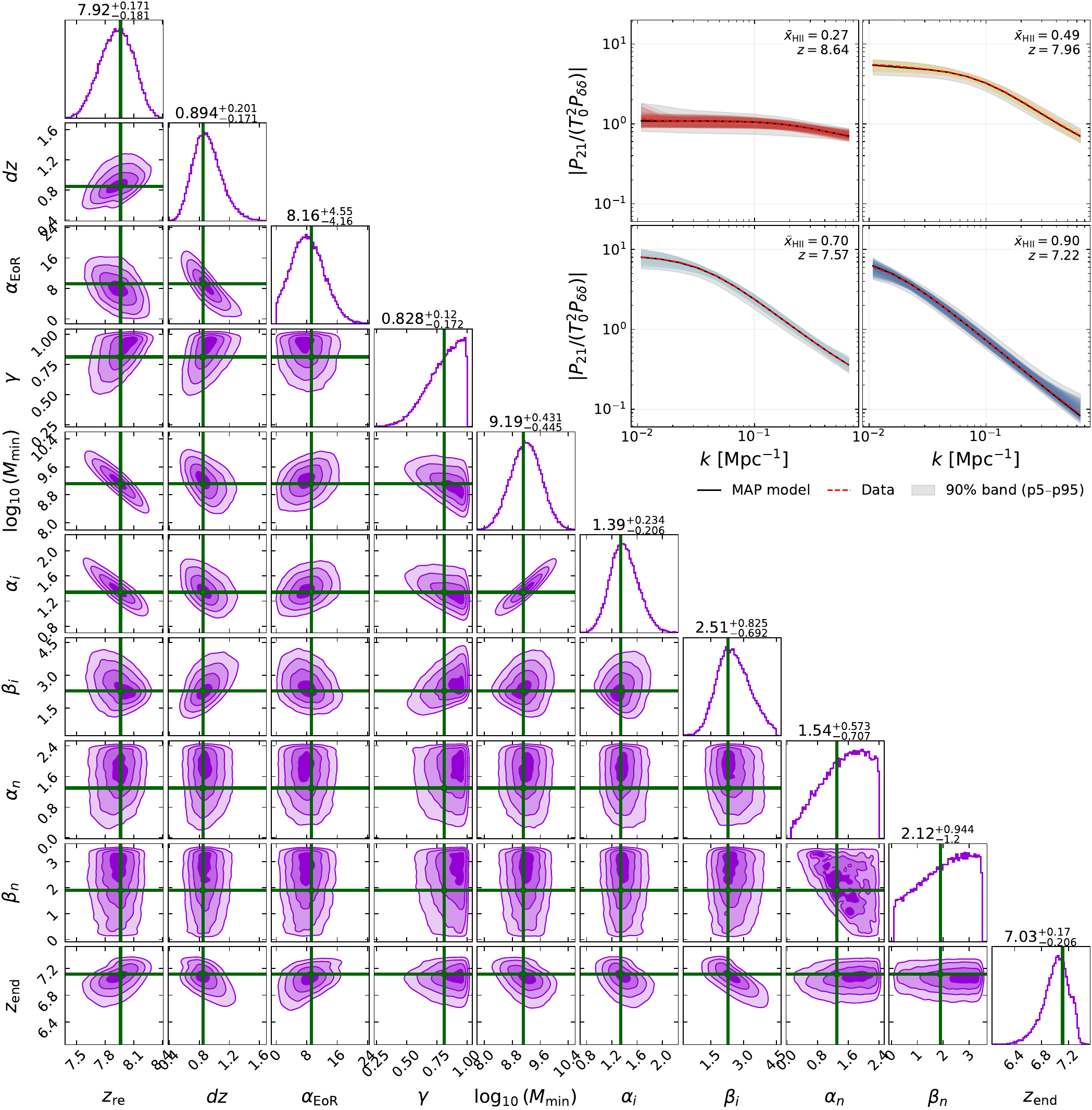}
    \caption{Posterior distribution from a forecast study generated from the best-fit results of the \texttt{C2R} simulation in Sec.~\ref{sec:results_c2ray} and comparable to the constraints in Fig.~\ref{fig:c2ray-posterior}. Correlations between the evolution parameters are also quantified by the Pearson $r$ coefficient, shown in Fig.~\ref{fig:forecast_correlation}.}
    \label{fig:forecast}
\end{figure*}
\begin{figure}
    \centering
    \includegraphics[width=.99\linewidth]{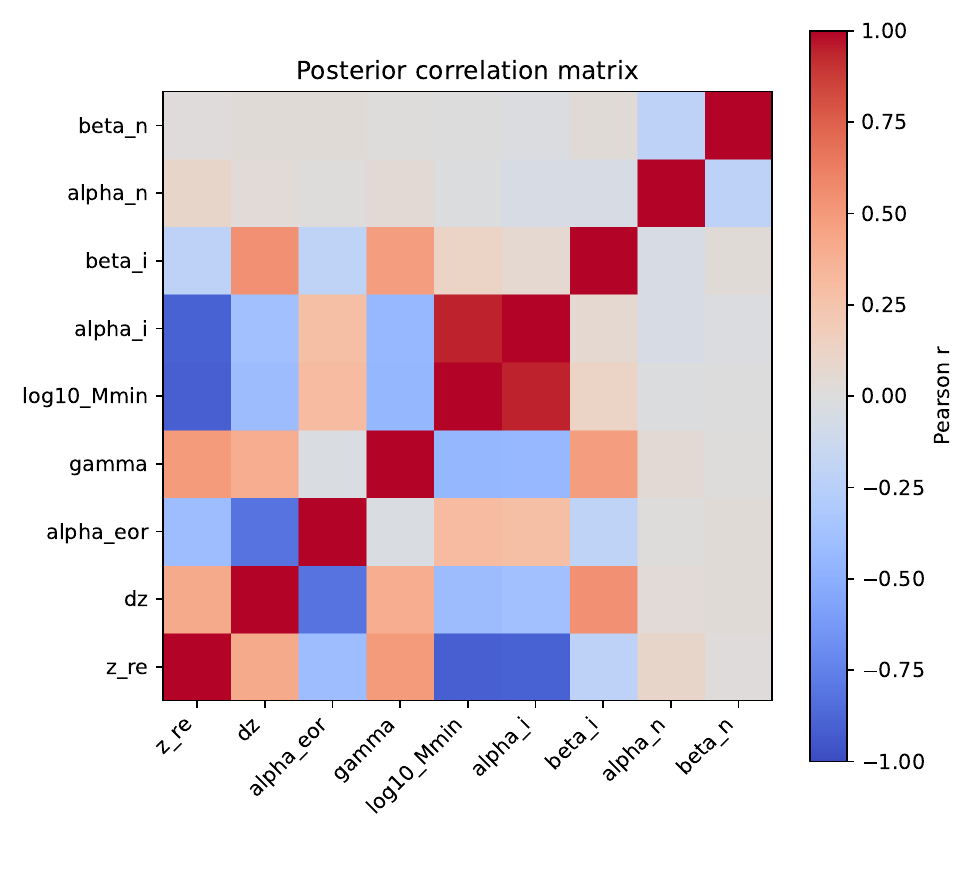}
    \caption{Pearson r coefficient calculated for the evolution parameters (outlined in Sec.~\ref{sec:IGM_params}), computed for the forecast model of \texttt{C2R}.}
    \label{fig:forecast_correlation}
\end{figure}
We apply the parametrisations in Sec.~\ref{sec:theory} to generate the volume and mass ionisation histories from Eqs.~\eqref{eq:reion_parametrisation} \&~\eqref{eq:xhii_m}, the evolution of ionised regions and neutral islands via Eqs.~\eqref{eq:Ri} \&~\eqref{eq:Rn} and halo bias evolution from Eq.~\eqref{eq:halo_bias}. These outputs are then combined to create the 21-cm power spectra data set using the analytic form in Eq.~\eqref{eq:p21_model}, including a random fractional error of 5 per cent across the EoR.

We run the sampler analysis ensuring an initial burn-in sample of 20 per cent of the original run with walker chains of the order $10^{6}$, reaching Gelman-Rubin statistic factors $\hat{R}-1<10^{-2}$ matching the initial set-up in Sec.~\ref{sec:results_c2ray}. We present the resulting posterior distribution in Fig.~\ref{fig:forecast} directly comparable to that of the \texttt{C2R} model in Fig.~\ref{fig:c2ray-posterior}. Both figures look almost identical in shape and correlation between the parameters, the small differences in the forecast posteriors further outline the discussion points in Sec.~\ref{sec:results_c2ray}. We examine the correlations between our parameters by quantifying them with the Pearson $r$ coefficient presented in Fig.~\ref{fig:forecast_correlation}, which measures the correlation and strength between two parameters. For the global ionisation parameters, we observe degeneracy between the ionisation rate parameter $\alpha_{\mathrm{EoR}}$ and the EoR duration, $dz$. More prominently, this is evident in the bottom-left part of Fig.~\ref{fig:forecast_correlation}, where the EoR rate and duration parameters are strongly anti-correlated. This is a sampling effect produced by the relation of the parameters seen in Eq.~\eqref{eq:reion_parametrisation}, where we can observe that a lower duration of the EoR can be overcompensated by a faster rate (approximately as $\log(\xa) \approx -\log(dz)\, \alpha_{\mathrm{EoR}}$). A strong positive correlation is also observed in the middle of Fig.~\ref{fig:forecast_correlation} between the powers of the average size of ionised bubbles at the midpoint of the EoR, $\alpha_{i}$ and the minimum mass of halos hosting ionising sources, $M_{\mathrm{min}}$. As outlined in Sec.~\ref{sec:results_c2ray}, this degeneracy arises because both IGM parameters $R_{i}$ and $b_{\mathrm{h}}$ directly influence the large-scale amplitude of the 21-cm power spectrum in Eq.~\eqref{eq:p21_model}. Both parameters have the most constraining power in the first half of cosmic reionization, yet the degeneracy can be naturally broken by tightening the constraints on $\alpha_{i}$ from observations.
\section{Discussion on the {\sc POLAR} 21-cm power spectrum decomposition}
\label{sec::polar_decomposition}
\begin{figure}
	\includegraphics[width=\columnwidth]{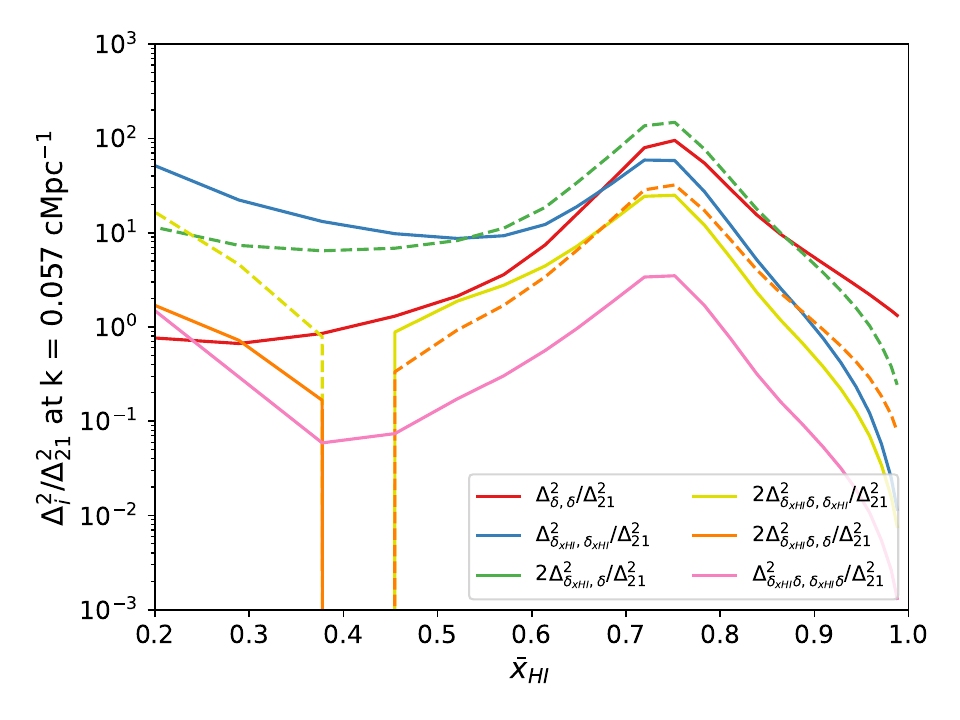}
    \caption{The fractional contribution from each power spectrum term from Eq.~\eqref{eq:Decomposition} at $k = 0.057~\mathrm{Mpc}^{-1}$ during the EoR. The solid lines represent the positive terms, while the dashed lines show the absolute values of the negative terms. Note that $\Delta^{2}_{\mathrm{a,b}}(k)=k^{3} P_{\mathrm{a,b}}(k)/(2\pi^{2})$.}
    \label{fig:polar_dec_terms}
\end{figure}
\begin{figure}
    \centering
    \includegraphics[width=1.\columnwidth]{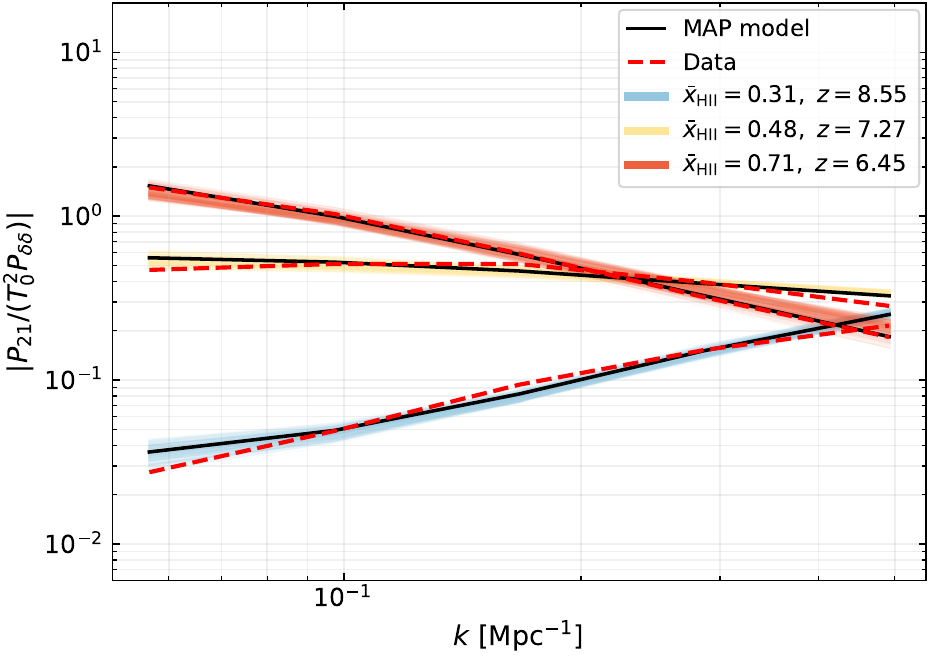}
    \caption{Ratio of the 21-cm power spectrum to the matter density power spectrum for the {\sc POLAR} simulation from Sec.~\ref{sec:Polar} as shown in Eq.~\eqref{eq:Decomposition} throughout the EoR, inferred from the posterior distributions shown in Fig.~\ref{fig:polar-posterior}. The true spectra are shown as a dashed red line, the best model as a black line, a selection of good $\chi^{2}$ models in the coloured regions, and the shaded regions show the sampled parameter space.}
    \label{fig:polar-p21-spectra}
\end{figure}
We consider the decreased low-$k$ power of the 21-cm ratio in Fig.~\ref{fig:polar-p21-spectra} and its effect on the constraint on the midpoint of reionization ($z_{\mathrm{re}}$) in Fig.~\ref{fig:polar-posterior}.
Both relate to the large-scale correlation between the density and ionisation fields, expressed through the higher-order terms in Eq.~\eqref{eq:Decomposition}. We can better understand this by examining Fig.~\ref{fig:polar_dec_terms}, where we plot each cross-term divided by the normalised 21-cm power against the global ionisation history to quantify each term's role in shaping the composite 21-cm spectra during the EoR. We note two distinct features, the peak of the decomposition occurs at $\xb \approx 0.75$, as well as a sign flip of the cross-order terms (seen in yellow and orange) at $\xb \approx 0.37$. Comparing the decomposition with results from {\sc C$^{2}$-Ray} and {\sc 21cmFAST} in fig.~4 of \citet{Georgiev2022} (see sec.~4 for a detailed discussion), the same features occur at $\xb \approx 0.90$ and $\xb \approx 0.5$, respectively. The former can be associated with the epoch of equilibration, a theoretical period in the early EoR where the distribution of ionised regions has grown sufficiently large around the highest density peaks, associated with their local matter over-densities, resulting in a drop in the large-scale 21-cm power spectrum \citep{Mellema2006b,Lidz2008}. The latter feature is often identified with the midpoint of reionization, during which the ionised and neutral IGM occupy the same volume of the simulation, after which the cross terms $P_{\delta_{\rho}\delta_{x_{\HI}},\delta_{\rho}}(k)$ \& $P_{\delta_{\rho}\delta_{x_{\HI}},\delta_{x_{\HI}}}(k)$ change sign, dominated by the fluctuations of the neutral regions located in low-density regions. 
The perceived delay of around 9-12 per cent in the decomposition's evolution is directly linked to the galaxy formation model used within the \texttt{PLR} model to assign ionising sources to dark matter halos. We can confirm this by analogy to the work of \citet{Iliev2012}, where the authors study the 21-cm power spectrum of HMACH-only reionization simulations against scenarios that include both HMACH and LMACH in their model, the latter halos being self-regulated by the role of negative feedback through Jeans mass filtering. Despite the different origins of the feedback mechanisms, radiative feedback in \citet{Iliev2012} and galactic feedback in \citet{Ma2026}, the resulting influence on the 21-cm power spectra is the same as in fig.~20 of \citet{Iliev2012}. In the right panel of the figure, one can clearly see the dip in power, associated with equilibration, for root mean square fluctuations of the 21-cm signal for models which do not include radiative feedback (e.g. 53Mpc$\_$g10.4$\_$0), which occurs at $\xb \approx 0.90$ whilst for the remaining models including various degrees of suppression the dip feature is shifted to values up to $\xb \approx 0.75$. This is the exact behaviour we observed and is therefore clearly linked to the role of suppression mechanisms on the LMACH sources in our simulation. Simply put, including LMACH sources allows a patchier ionisation morphology, rich in small ionised regions, whose feedback mechanisms suppress their growth as the EoR progresses. Hence, our simulation box has more partially ionised cells that have either never been fully ionised or, due to the lack of ionising sources, are slowly becoming neutral through radiative recombination. This is in contrast to HMACH-only simulations, where partially ionised cells are fewer and often mainly associated with regions where I-fronts are currently propagating and thus the cells are not yet in ionisation equilibrium \citep{Zahn2011}. The overall effect of this is that for the models with suppression like \texttt{PLR}, the global ionisation fraction $\xa$ is higher than that of models such as \texttt{C2R} and \texttt{21F}. In fact, we can test this for the \texttt{PLR} model by applying a threshold such that any cell above $x_{\mathrm{th}}>0.5$ is set to unity, and vice versa. The resulting threshold shifts the equilibration global neutral fraction $\xb \approx 0.75 \rightarrow {\bar{x}_{\rm \ion{H}{I} th>0.5 }} \approx 0.87$ and cross-power spectrum sign flip at $\xb \approx 0.37 \rightarrow {\bar{x}_{\rm \ion{H}{I}th>0.5}} \approx 0.46$, in agreement with our expectations. 

How do we interpret these results in the context of the IGM framework? On the one hand, Fig.~\ref{fig:polar-p21-spectra} shows that sign flips in the cross-order term play a small role in the late-time 21-cm ratio at $\xa = 0.71$. This is corroborated by the decomposition results in Fig.~\ref{fig:polar_dec_terms}, where the cross-order terms are clearly subdominant in the overall 21-cm power-spectrum amplitude compared to the auto- and cross-correlation power spectra of the ionisation field. In contrast, the equilibration epoch completely dominates the 21-cm amplitude at $\xa = 0.31$ in Fig.~\ref{fig:polar-p21-spectra}, and its large-scale $k$-dependence can be associated with the equilibration process. The partially ionised cells associated with suppressed sources both delay equilibration in the EoR and couple the ionisation and density fields more strongly on larger scales. This increases the amplitude of the cross-correlation power spectrum $P_{\delta_{\rho},\delta_{x_{\HI}}}(k)$, which remains a dominant term in Fig.~\ref{fig:polar_dec_terms} up to the midpoint of the EoR. This results in a drop in low-$k$ power, unlike HMACH source models, where the autocorrelation power dominates. In turn, the IGM framework captures the decrease in the low-$k$ 21-cm ratio by reducing the minimum halo mass and/or the reionization duration $dz$ (as previously highlighted in Fig.~\ref{fig:polar-posterior}).

\bsp	
\label{lastpage}
\end{document}